\documentclass[useAMS,usenatbib,usegraphicx]{mn2e}
\usepackage{color}

\title[A comparison of period finding algorithms]{A comparison of period finding algorithms}
\author[M. J. Graham et al.]{Matthew~J.~Graham,$^1$\thanks{E-mail:mjg@caltech.edu} Andrew~J.~Drake,$^1$
S.~G.~Djorgovski,$^1$   
\newauthor
Ashish~A.~Mahabal,$^1$ Ciro~Donalek$^1$, Victor~Duan$^1$, and Alison~Maher$^1$\\
$^{1}$California Institute of Technology, 1200 E. California Blvd, Pasadena, CA 91125, USA}
\begin{document}

\date{Accepted . Received ; in original form}

\pagerange{\pageref{firstpage}--\pageref{lastpage}} \pubyear{2011}

\maketitle

\label{firstpage}

\begin{abstract}
This paper presents a comparison of popular period finding algorithms applied to the light curves of variable stars from the Catalina Real-time Transient Survey (CRTS), MACHO and ASAS data sets. We analyze the accuracy of the methods against magnitude, sampling rates, quoted period, quality measures (signal-to-noise and number of observations), variability, and object classes. We find that measure of dispersion-based techniques --  analysis-of-variance with harmonics and conditional entropy -- consistently give the best results but there are clear dependencies on object class and light curve quality. Period aliasing and identifying a period harmonic also remain significant issues. We consider the performance of the algorithms and show that a new conditional entropy-based algorithm is the most optimal in terms of completeness and speed. We also consider a simple ensemble approach and find that it performs no better than individual algorithms. 
\end{abstract}


\section{Introduction}
The last decade has seen the emergence of large collections of time series data from searches for microlensing, e.g., MACHO (\cite{macho}), OGLE (\cite{ogle}), and exoplanets, e.g., CoRoT (\cite{corot}), Kepler (\cite{kepler}), as well as legacy variability collections, e.g. ASAS (\cite{asas}). For the first time, these are amenable to statistical and machine learning-based analyses, particularly for classification and outlier detection,
e.g., \cite{deboss07}, \cite{shin}, \cite{dubath}, \cite{richards11}, with an eye to the new generation of synoptic sky surveys, e.g., CRTS (\cite{crts}), PTF (\cite{ptf}), Pan-STaRRs (\cite{panstarrs}), and LSST (\cite{lsst}), which will increase the amount of available data by several orders of magnitude. These surveys are also not unique to optical wavelengths with efforts underway across the electromagnetic spectrum -- LOFAR and SKA and its pathfinder precursor projects in the radio, IR, X-ray -- as well as in the more exotic regimes of particle astrophysics (neutrino) and gravitational waves (LIGO). Although many different approaches have been attempted, they all follow the same basic pattern: characterization, categorization, and classification.

Time series vary widely in their temporal coverage, sampling rates and regularity, number of points and error bars, making a very disparate data set. Comparing raw light curves\footnote{Note that we use the terms ``time series'' and ``light curve'' interchangeably in this paper.} is therefore difficult; rather a representation of each light curve in a given data collection in terms of a feature set is required for any analysis. There is no standardized set -- somewhere around 100 different features\footnote{A feature is defined as an individual measurable heuristic property of an object that can be used to characterize it.} have been used or suggested in the literature for characterizing time series, e.g., moments, flux and shape ratios, variability indices, etc. -- but many of them rely on a derived period for an object, even when it does not necessarily display any periodic behaviour. \cite{dubath} show a $\sim$11\% misclassification error rate for non-eclipsing variable stars with an incorrect period. \cite{richards11} also estimate that periodic feature routines account for 75\% of the computing time used in their time series characterization.

This irregularity means that astronomical time series data does not lend itself to the standard Fourier-based analysis techniques that are found in general statistics literature. Consequently, there is a long history in period finding algorithms with common ones based upon discrete Fourier transform (\cite{deeming}),  or a least-squares approximation to it (LS; \cite{lomb, scargle}), string length (\cite{dworetsky}, phase dispersion (PDM; \cite{jurkevich, stellingwerf}), and analysis of variance (AOV) methods (\cite{aov89, aovmhw}), as well as a host of others (e.g., \cite{correntropy, lasso}).   

Obviously with so many different methods, the question arises as to which one is the best, if any. 
\cite{heckmm} used numerical simulations to compare discrete Fourier transforms, string length and phase dispersion methods and found that none of them was superior to the others. \cite{schwarzenberg99} compared model function and phase binning methods using hypothesis-testing theory to evaluate their relative sensitivity to different kinds of signals. He found that the methods using smooth model functions, such as LS, are more sensitive than those using the step function, i.e., phase binning, and that sensitivity increases for models that more closely fit features in the signal with the orthogonal multiharmonic AOV method (AOVMHW; \cite{aovmhw}) being optimal. 

He also found that a number of methods relying on phase binning are equivalent for the same number of bins. Similarly, \cite{swingler89} argues that PDM methods should be regarded simply as approximations to the Fourier method (LS) and that the latter should be viewed as the periodogram technique of choice.
\cite{distefano} have compared discrete Fourier, LS and PDM techniques for recovering the rotation periods of solar-like stars from irregular time sampling of {\it Gaia} using synthetic time series. They find that LS is the most efficient method with at best a recovery rate of $\sim 60$\%.

\cite{dubath} report that a single method can lead to a recovery fraction of around 80\% but do not specify to what degree of accuracy. They also suggest that an ideal combination of all methods could potentially raise that value to close to 100\% but cannot identify the automated strategy for predicting which method leads to the correct period for a specific light curve. However, they propose, as a first step, a combination of unweighted and weighted Lomb-Scargle, depending on the skewness of a source's magnitude distribution.

In this work, we present a detailed comparison of the most commonly used period finding algorithms and their efficiencies against observable parameters. This is the first survey using real rather than simulated data (so with noise, gaps, etc.) to consider both a wide range of variable stellar classes and light curves generated by different sampling strategies. It is hoped that we can identify the most effective algorithm with a particular view to the next generation of survey projects which require automated and efficient period finding methods. 

The paper is structured as follows: in section 2, we describe the data sets that form the basis of our analysis whilst in section 3, we present the algorithms that we are considering here. We analyze and discuss our results in sections 4 and 5 and present our conclusions in section 6. We also have provided implementation details about OpenCL-based versions of the Lomb-Scargle and generalized Lomb-Scargle algorithms in an appendix.

\section{Data sets}
\begin{figure}
\caption{This shows the V-band magnitude distribution of the three data sets considered in this paper: ASAS (red), MACHO (black) and CRTS (blue).}
\label{magst}
\includegraphics[width=3.2in]{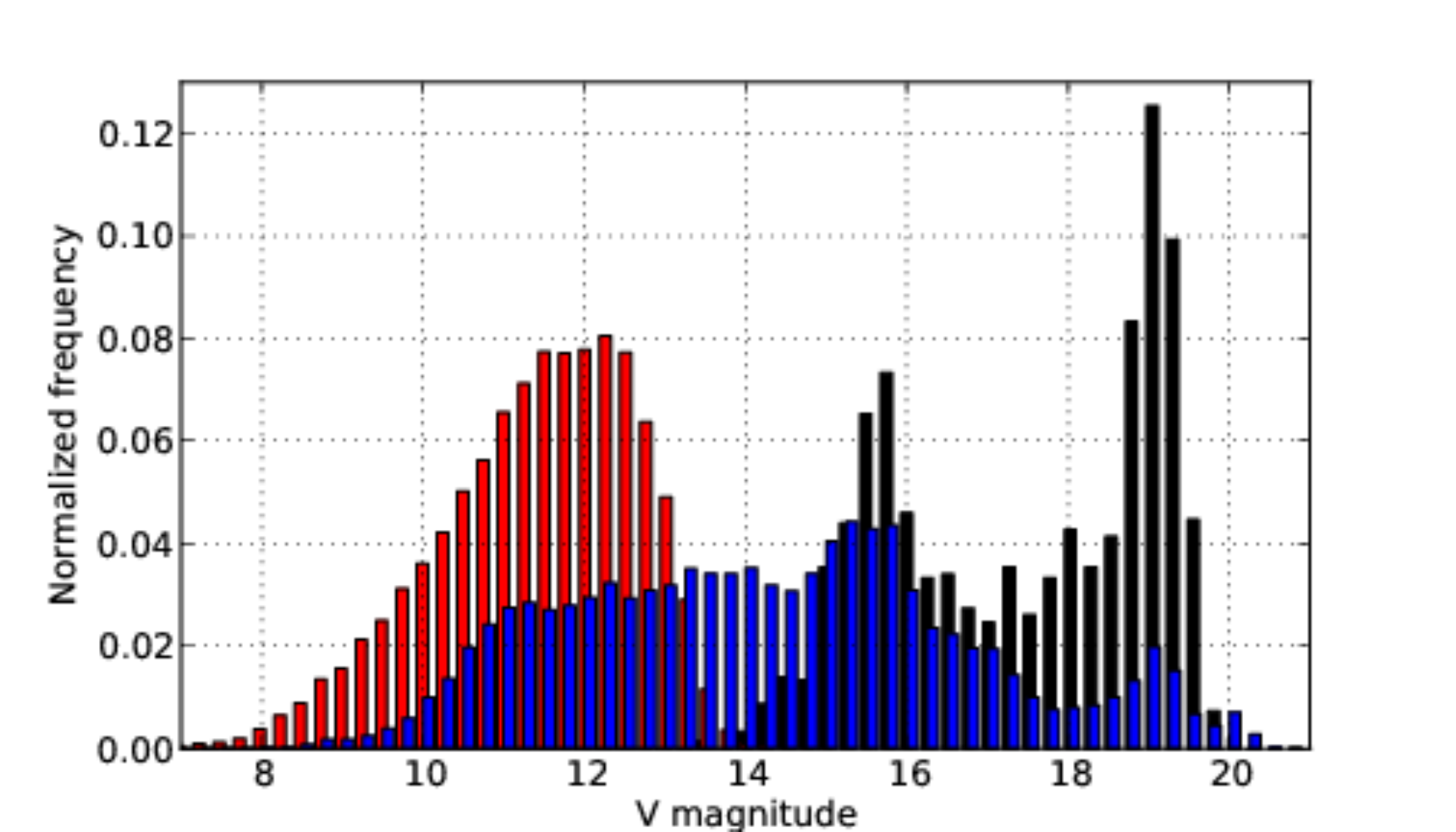}
\end{figure}

\begin{figure}
\caption{This shows the distribution of observations per light curve for the three data sets considered in this paper: ASAS (red), MACHO (black) and CRTS (blue).}
\label{counts}
\includegraphics[width=3.2in]{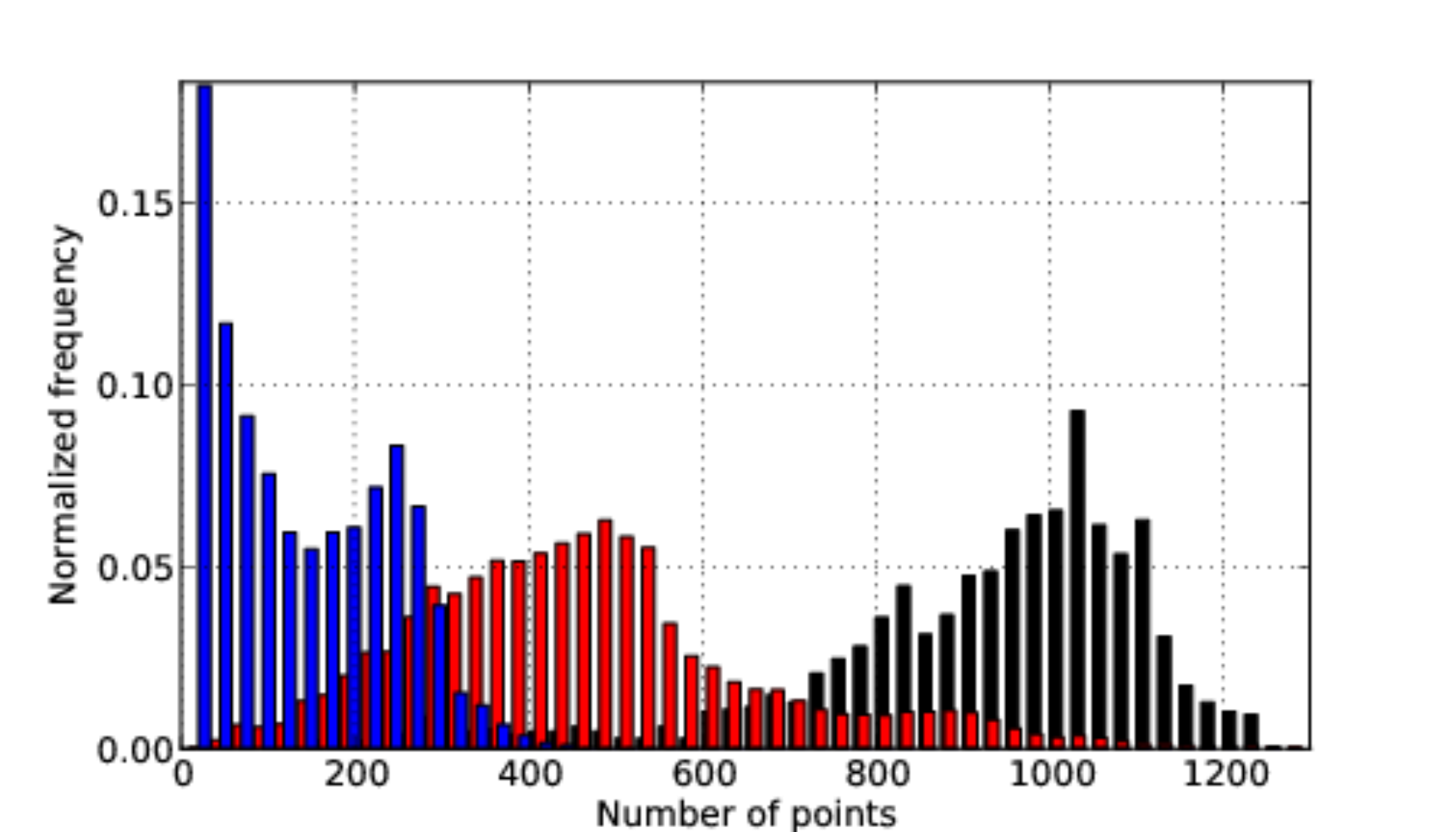}
\end{figure}

In this analysis, we consider three sets of light curves, drawn from different surveys -- CRTS, ASAS, MACHO, which we take to be representative of the bulk of ground-based light curve data sets currently available and characteristic of future large samples such as LSST. Together these span a magnitude range of $\sim4 \leq V \leq 21$ (see Fig.~\ref{magst}) and a sampling of up to $\sim1800$ observations (see Fig.~\ref{counts}) over a baseline of up to $\sim8$ years. Details of the three are summarized in Table~\ref{datasumm}. All observation times are converted to Heliocentric Julian Date (HJD) from MJD.

\begin{table*}
\caption{This summarizes the three data sets used in this analysis.}
\label{datasumm}
\begin{tabular}{lccccc}
\hline
Data set & No. of sources & \multicolumn{4}{c}{Median values} \\
 & & Magnitude & Observations & Baseline (d) & Period (d) \\
\hline
ASAS & 50124 & 11.59 & 456 & 2645 & 44.51 \\
CRTS & 15522 & 14.35 & 105 & 2182 & 0.59 \\
MACHO & 1500 & 17.82 & 966 & 2721 & 1.74 \\
\hline
\end{tabular}
\end{table*}

\subsection{Catalina Real-time Transient Survey (CRTS)}
The Catalina Real-time Transient Survey (\cite{crts}) is the largest open time domain survey currently operating, covering $\sim$33000 deg$^{2}$ between $-75^{\circ} < {\mathrm Dec} < 75^{\circ}$ (except for within $\sim$10 -- 15$^{\circ}$ of the Galactic plane). It leverages the data streams from 3 telescopes used in a search for near-Earth objects, operated by the Lunar and Planetary Laboratory at U. of Arizona,  with 4 exposures per visit, separated by 10 min, reaching to V $\sim$ 19 to 21.5 mag (depending on telescope), over 21 nights per lunation. All data are automatically processed in real-time, and optical transients are immediately distributed using a variety of electronic mechanisms\footnote{http://www.skyalert.org}. Light curves of several hundred million objects are available\footnote{http://crts.caltech.edu} with an average of $\sim$250 observations over a 8-year baseline. 

\begin{figure*}
\centering
\caption{This shows the sky distribution in galactic coordinates of ASAS (left) and CRTS (right) sources. MACHO sources are localized to the LMC and not shown. The sources are color-coded according to their broad variable class: eruptive (black), pulsating (red), rotation (maroon), cataclysmic (yellow), eclipsing (cyan), X-ray (blue) and other (green). Larger symbols are employed for the less populous classes.}
\label{skypos}
\includegraphics[width=6.4in]{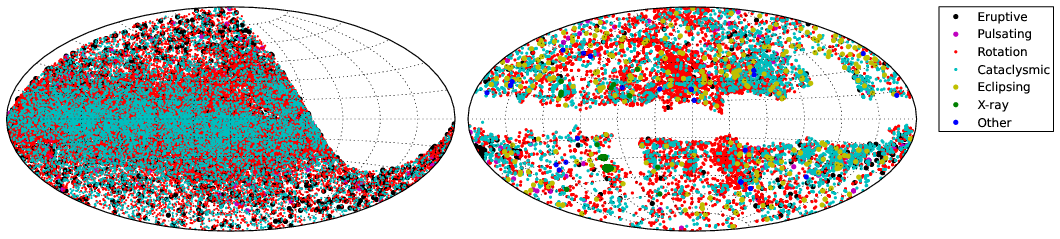}
\end{figure*}

To get a sample of as wide a range of variable sources as possible, all objects in SIMBAD and the AAVSO Variable Star Index (VSX, \cite{vsx}) with a recorded period were selected, giving 146655 sources (71\% of the total combined number). Light curves were then extracted for those which had been observed by CRTS. Since very many of the initial data set lie in the galactic plane and CRTS explicitly avoids the galactic plane, this brought the number of sources covered to a manageable 15522 (see Fig.~\ref{skypos} for the distribution of the sources on the sky). The poorer sampling near the plane also explains why the median number of observations for this data set is 105.

All the light curves were inspected by three of us to verify (via phased light curves) the quoted period against fit periods from two other methods: AOV and conditional entropy (CE, \cite{graham13}). When there was no consensus opinion, the quoted period was used. 

\subsection{ASAS}

The ASAS Catalog of Variable Stars (ACVS; \cite{acvs}) represents one of the largest collections of light curves of variable stars available, covering a range of 11 science classes (albeit predominantly MISC) and with good consistent data.  The sky distribution with the limit $\delta <  +28$ is shown in Fig.~\ref{skypos}. \cite{macc12} (MACC) have applied probabilistic classifiers to ACVS light curves to create a 28-class machine-learned catalog of 50124 sources. We have followed a similar prescription to MACC to construct our data set of ACVS light curves:  the data for individual objects are retrieved from the ACVS website (ACVS 1.1\footnote{http://www.astrouw.edu.pl/asas/}) and those epochs with a quality {\tt GRADE=D} or quality {\tt GRADE=C} when {\tt MAG=29.999} excluded, corresponding to a non-detection. This gives a median of 456 usable epochs of V-band observations covering 2644.92 days.

ASAS provides five aperture measurements using diameters ranging from 2 pixels (30'') to 6 pixels (90'') and describes a basic algorithm for choosing which aperture to use for an object given its average magnitude (\cite{acvs}). MACC constructed a simple classifier to determine the optimal aperture to use for each object which we have followed: using 2 pixels for $V > 12.25$, 3 pixels for $11.675 < V < 12.25$, 4 pixels for $10.675 < V < 11.675$, and 6 pixels for $V < 10.675$.

ACVS periods were determined using an AOV algorithm and confirmed visually. MACC has also determined a period for each object using a generalized Lomb-Scargle-based algorithm (\cite{gls}) with corrections for eclipsing and aliased periods (see Sec.~3.2). The quoted agreement between the two is 77.2\% (exactly matching) for the 12008 objects  which ACVS confidently classified into a single periodic class (see Sec.~4.1 for a discussion of this). We have inspected a representative sample of the light curves and consider the ACVS period to be the true value.

\subsection{MACHO}

The MACHO survey (\cite{macho}) was designed to search for gravitational microlensing events in the Magellanic Clouds and the Galactic Bulge and more than 20 million stars were observed, making it an important resource for variable star studies. A ``gold standard'' data set of light curves has been produced from the MACHO survey by the Harvard Time Series Center\footnote{http://timemachine.iic.harvard.edu}, consisting of approximately 500 each of RR Lyrae, eclipsing binaries and Cepheids respectively covering the LMC ($75^{\circ} < \mathrm{RA} < 85^{\circ}$, $-71^{\circ} < \mathrm{Dec} < -67^{\circ}$). Although MACHO data normally consists of blue and red channel data for each stellar object, only the blue channel (V-band equivalent) have been used here. The median time span of the data is 2720.88 days. This data set has also been used in two correntropy-based (generalized correlation) approaches for estimating periods in non-uniformly sampled time series: \cite{misha11} employs slots (intervals) (\cite{misha11}) to determine the statistic of interest whilst \cite{ckp} uses a kernel.

\subsection{Variable classes}

Objects in the CRTS data set have been labelled with classes drawn from VSX\footnote{http://www.aavso.org/vsx/help/VariableStarTypeDesignationsInVSX.pdf} which is itself based on 
the General Catalog of Variable Stars (GCVS; \cite{gcvs}) classification scheme (a maximum crossmatch distance of 3'' was used). This is a relatively detailed system that covers most types of variable stellar phenomena. Objects fall into one of seven broad classes reflecting both extrinsic and intrinsic phenomena: eruptive, pulsating, rotating, cataclysmic, eclipsing, X-ray and other. For convenience, we have converted the VSX codes into a hierarchical coding scheme: for example, the eclipsing class is P.5, an eclipsing binary is P.5.1 and a $\beta$ Lyrae-type eclipsing binary is P.5.1.2.
 
For ASAS data, we use the MACC classifications from ASAS\_CATALOG\_CLASS\_V3.0\footnote{http://www.bigmacc.info/}. MACC employs a twenty eight term scheme taken from \cite{deboss07} with the addition of SX Phe and splitting T Tauri into two subclasses: classic T Tauri and weak-line T Tauri. Six of these classes do not have an equivalent in the VSX/GCVS scheme and so we add them to ours - they are: long secondary period red giants (LSP), small amplitude red giants split according to \cite{wood07} (SARG\_A, SARG\_B), red supergiants (RSG), chemically peculiar (ChemPec), and Herbig AE/BE (HAEBE). Note that only 8572 out of the 50124 ASAS sources have a MACC classification with a probability of 90\% or higher.

\begin{figure}
\caption{This shows the distribution of MACHO light curves in the median absolute deviation (from the median) - period plane. The different colors denote different MACHO classes of object: blue (EB), red (RRAB), cyan (Cepheid fundamental), green (RRC), purple (RRE),  black (Cepheid first overtone). The crosses indicate objects for which there is no MACHO classification in the literature and one must be imputed by a nearest-neighbour classifier.}
\label{tsc_class}
\includegraphics[width=3.37in]{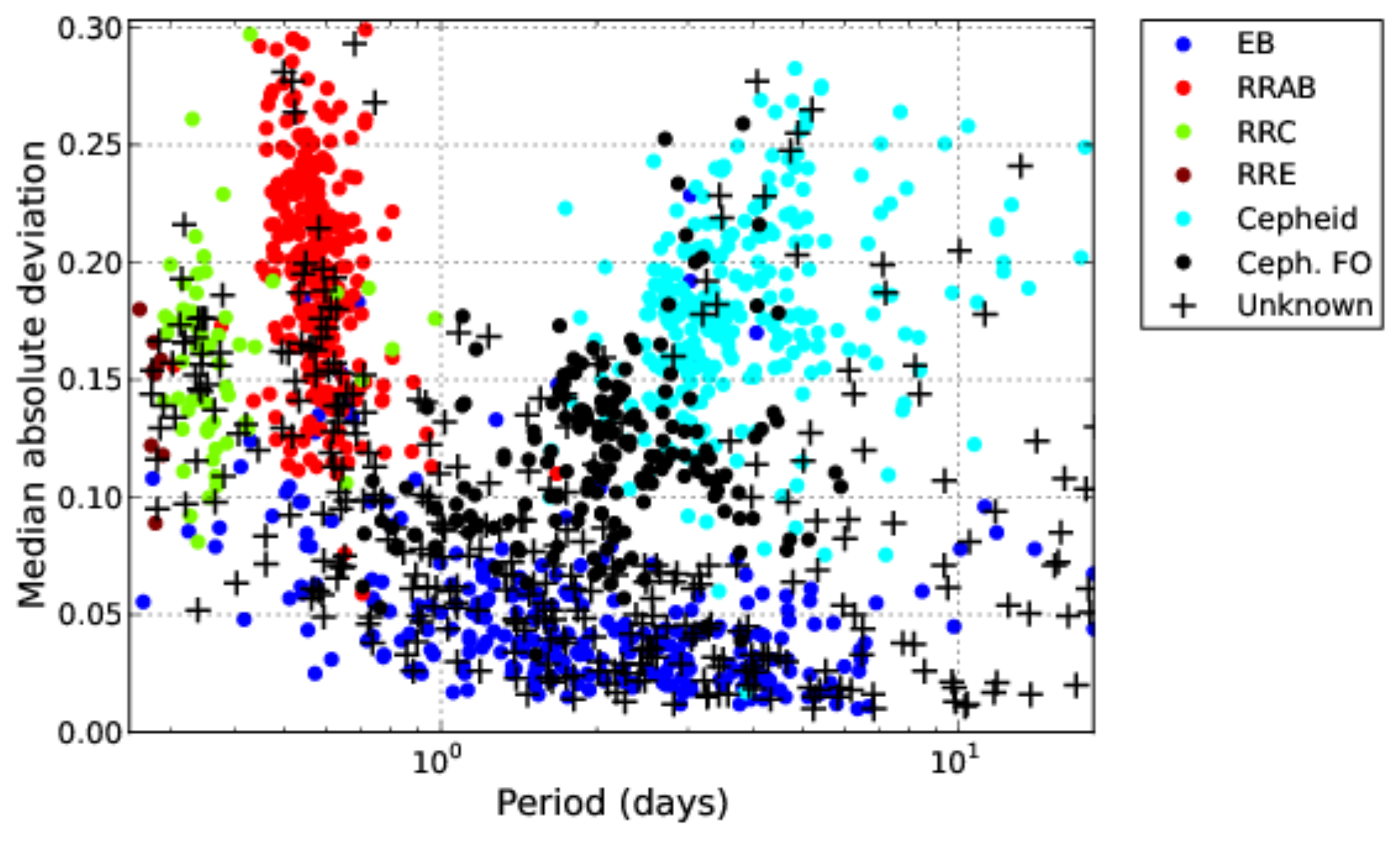}
\end{figure}

The MACHO data initially consisted of just three classes: RR Lyrae, eclipsing binaries and Cepheids. However, additional data for these objects (\cite{macho}) gave finer-grained classifications for 1139 stars based on an automated statistical analysis of its photometry over time. A plot of the median absolute deviation from the median (MAD) of the light curves of these objects vs. their quoted period can be useful for discriminating between different classes (see Fig.~\ref{tsc_class}). A nearest-neighbour classifier in the MAD-period plane was then used to impute classes to the remaining 361 objects without finer-grained classifications. These were checked with SIMBAD: of the 305 objects with a SIMBAD classification but not a fine-grained one, $\sim$91\% agreed with their imputed class. This is the same level of accuracy as the MACC classifications.

\begin{figure*}
\caption{This shows the distribution of quoted periods in days for the three data sets considered in this paper: ASAS (red), MACHO (black) and CRTS (blue).}
\label{periods}
\includegraphics[width=7.5in]{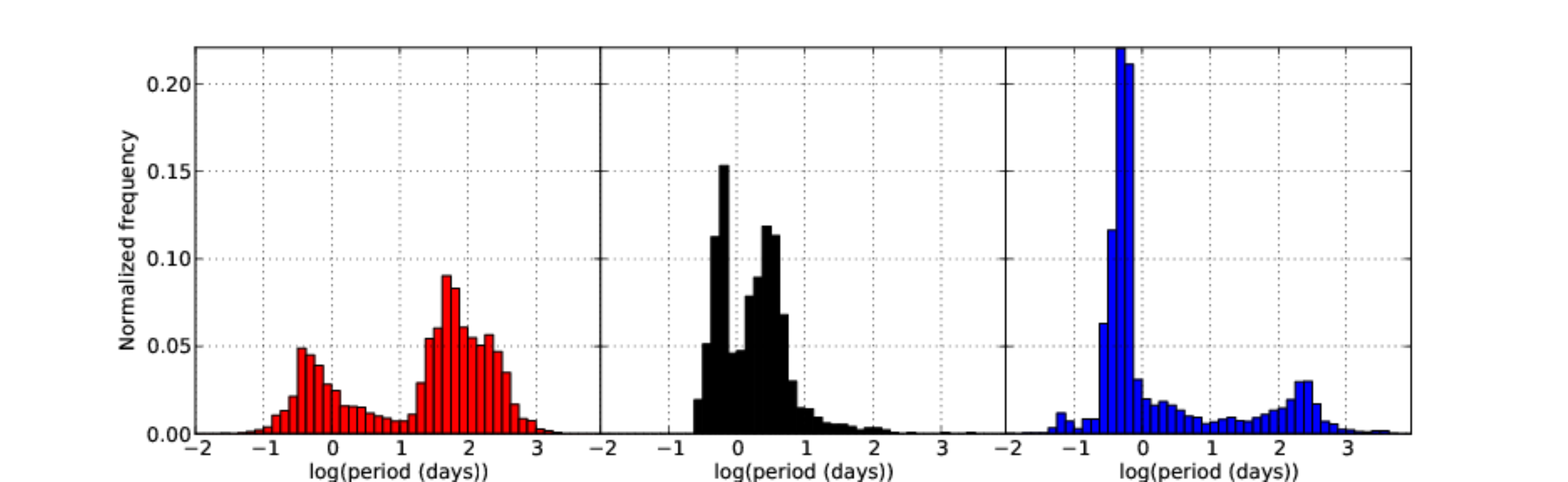}
\end{figure*}

Table~\ref{types} gives the relative numbers of objects per class in the three data sets considered in this paper. The distribution of periods over the three data sets is shown in Fig.~\ref{periods}. As mentioned above, these have all been visually confirmed (either by us or other authors) and so this is the true distribution with no contamination from aliased periods. The peaks at log(period) $\sim -0.4$ in the three distributions are from RR Lyrae and eclipsing binary objects. Similarly, the peak at log(period) $\sim 1.5$ in the ASAS data is from small amplitude red giants, the peak at log(period) $\sim 0.5$ in the MACHO data from Cepheids, and the peak at log(period) $\sim 2.5$ in the CRTS data from Mira variables respectively.

\begin{table*}
\caption{The relative numbers of each class of variable stellar object used in this analysis. Only those classes which have instances are included. The codes in parentheses are the GCVS or VSX code for this type of variable or, if marked with an asterisk, the code used in MACC. The method given is the most reliable method for finding periods for this class (see Sec.~\ref{reliability} for details) with an asterisk indicating that less than 10 periods were recovered by it. A dash denotes that no method recovered an accurate period for the class. When two methods are given, they both recovered the same number of periods accurately.}
\label{types}
\begin{tabular}{@{}llllll}
\hline
Class & Label & CRTS & MACC & MACHO & Method \\
\hline
\multicolumn{3}{@{}l}{{\bf Eruptive}} \\ 
Be variable & P.1.2.1 (BE) & - & 337 & - & GLS\\
Poorly studied irreg. var. of inter. to late spectral type & P.1.3.2 (IB) & 43 & - & - & AOVMHW*\\
Orion variable & P.1.3.3 (IN) & 5 & - & - & LS*\\
Orion variable - early spectral type & P.1.3.3.1 (INA) & 85 & - &  - & GLS*\\
T Tauri & P.1.3.3.3 (INT, IT) & 28 & 5 &  - & AOVMHW* \\
Weak-line T Tauri & P.1.3.3.3.2 (WSST) & - & 2239 & - & GLS \\
Rapid Orion variable of intemediate to late spectral type & P.1.3.3.6 (INSB)  & 2 & - & - & -\\
Rapid T Tauri & P.1.3.3.7 (INST)  & 5 & - & - & AOV*\\
Rapid irregular variable without nebula & P.1.3.4 (IS) & 3 & -& - & AOVMHW*\\
R Cor Bor type variable & P.1.4 (RCB) &- & 53 & - & GLS*\\
RS Can Ven type variable& P.1.5 (RS) & 178 & 263 & - & AOVMHW\\
S Dor type variable & P.1.6 (SDOR) &- & 1 &- & -\\
UV Ceti type variable & P.1.7 (UV) & 4 &- & - & -\\
Flaring Orion variable of spectral type Ke - Me & P.1.7.1 (UVN) & 1 & -& - &-\\
Young Stellar Object & P.1.9 (YSO) & 4 & -&- & -\\
Herbig AE/BE & P.1.10 (HAEBE$^{\ast}$) & - & 111 & - & CE* \\ 
Red supergiant & P.1.11 (RSG$^{\ast}$) & - & 827 & - & GLS \\
\hline
\multicolumn{3}{@{}l}{{\bf Pulsating}} \\ 
General pulsating variable & P.2 (PULS) & 58 & -& - & GLS\\
Beta Cephei type variable & P.2.2 (BCEP) & 3& 259& -& AOVMHW\\ 
Cepheid & P.2.3 (CEP) & 25 & 568& 287 & CE\\
Multimode Cepheid & P.2.3.1 (CEP(B)) & &202 & 230 & CE\\
W Vir type variable& P.2.4 (CW) &2 &- &-  & AOV\\
W Vir type variable with period longer than 8 days & P.2.4.1 (CWA) &26 & -& - & CE\\
W Vir type variable with period shorter than 8 days & P.2.4.2 (CWB) & 30& -&- &AOVMHW\\
Classical Cepheid (Delta Cep)& P.2.5 (DCEP) & 53&- &- &AOVMHW*\\
D Cep type variable with light amp. and symmetrical LC & P.2.5.1 (DCEPS) &2 &- &- & CE*\\
Delta Scuti type variable & P.2.6 (DSCT) & 92& 1527 &- & FC\\
Low amplitude Delta Scuti type variable & P.2.6.1 (DSCTC) &5 &- &- & AOVMHW*\\
High amplitude Delta Scuti type variable & P.2.6.2 (HADS) & 95 &- &- & LS\\
Slow irregular variable of late spectral type& P.2.7.1 (LB) & 7&- & - & AOVMHW/FC*\\
Mira type variable & P.2.8 (M) & 979& 3086 & -& GLS\\
RR Lyrae type variable & P.2.10 (RR) & 1957 &- & 11 & CE\\
RR Lyrae type variable with fundamental mode & P.2.10.1 (RRAB) & 4518 & 1460 & 343& CE\\
RR Lyrae type variable with fundamental overtone & P.2.10.3 (RRC) & 692 & 476& 91& AOVMHW\\
RR Lyrae type variable with double mode & P.2.10.4 (RRD) & 137 &130 & 15 & GLS\\
RV Tauri type variable & P.2.11 (RV)  &7 & 452& - & CE\\
RV Tauri type variable that does not vary in mean mag. & P.2.11.1 (RVA) & 18&- &- & AOVMHW*\\
RV Tauri type variable that varies periodically in mean mag. & P.2.11.2 (RVB) & 1& -& - & STR*\\
Semiregular variable & P.2.12 (SR) & 436& 9982 &- & GLS\\
Semiregular late-type giant with persistent periodicity  & P.2.12.1 (SRA) &142 &- &- & AOVMHW \\
Semiregular late-type giant with poorly defined periodicity & P.2.12.2 (SRB) &168 & -& - & AOVMHW\\
Semiregular late-type supergiant & P.2.12.3 (SRC) &1 &- &-& - \\
Semiregular variable giant/supergiant & P.2.12.4 (SRD) &16 &- & -& CE*\\
Semiregular variable & P.2.12.5 (SRS) &2 & -& -&-\\
SX Phe type variable & P.2.13 (SPXHE) &35 & 12 & -& FC\\
ZZ Ceti type variable & P.2.14 (ZZ) &2 &- &- & -\\
Anomalous Cepheid type variable & P.2.17 (BLBOO) &13 & -& - & AOVMHW*\\
G Dor type variable & P.2.18 (GDOR) & 2 &- &- & CE*\\
Small amplitude red giants - type A  & P.2.19.1  (SARG\_A$^{\ast}$) & -& 3974 & - & GLS\\ 
Small amplitude red giants - type B  & P.2.19.2 (SARG\_B$^{\ast}$) & -& 7820 & - & GLS \\ 
Long secondary period red giants & P.2.20 (LSP$^{\ast}$) & - & 5096 & - & GLS\\
\hline
\multicolumn{3}{@{}l}{{\bf Rotating}} \\ 
Alpha2 Can Ven type variable & P.3.1 (ACV) & 1&- &- & AOVMHW*\\
BY Draconis type variable & P.3.2 (BY) & 89 & -& -& AOVMHW \\
Ellipsoidal variable & P.3.3 (ELL) & 18 & 2 & - &AOVMHW*\\
Chemically peculiar & P.3.7 (ChemPec$^{\ast}$) & - & 345 & -  & GLS \\
\hline
\end{tabular}
\end{table*}

\begin{table*}
\contcaption{}
\begin{tabular}{@{}llllll}
\hline
Class & Label & CRTS & MACC & MACHO & Method \\
\hline
\multicolumn{3}{@{}l}{{\bf Cataclysmic}} \\ 
Cataclysmic variable & P.4 (CV) & 17 & -& -& CE*\\
Fast novae & P.4.1.1 (NA) & 4&- & - &-\\
Slow novae & P.4.1.2 (NB) &6 & -& -& PDM*\\
Novalike variable & P.4.1.4 (NL) & 57&- & -&AOVMHW*\\
Recurrent novae & P.4.1.5 (NR) & 5& -& -&AOVMHW*\\
U Gem type variable & P.4.3 (UG) & 88&- &- &PDM2*\\
SS Cyg type variable & P.4.3.1 (UGSS) & 21& -& -&PDM/CE*\\
SU U Ma type variable & P.4.3.2 (UGSU) &140 & -&- &AOVMHW\\
Z Cam type variable & P.4.3.3 (UGZ) & 22&- &- &AOVMHW*\\
WZ Sag type variable & P.4.3.4 (UGWZ) & 19 &- & -&-\\
Z Andr type variable & P.4.4 (ZAND) & 1&- &- &-\\
 & P.4.5 () & 6 & & & \\
\hline
\multicolumn{3}{@{}l}{{\bf Eclipsing}} \\ 
Eclipsing binary system & P.5.1 (E) & 109& -& 522 & CE\\
Beta Persei (Algol) type system & P.5.1.1 (EA) & 1025 & 2855 & -& STR\\
Beta Lyrae type system & P.5.1.2 (EB) & 866& 1963& -& CE\\
W U Ma type system & P.5.1.3 (EW) &1479 & 6025 & - & AOVMHW\\
Contact system & P.5.8 (K) & 20 & -& - & CE\\
Semi-detached system & P.5.9 (SD) & 1 &- & -& STR*\\
AM Her type system & P.5.10 (AM) &76 &- &- & CE*\\
Close binary system with strong reflection & P.5.11 (R) & 9&- &- & CE*\\
Planet eclipsing system & P.5.12 (EP) & 2& -& -&-\\
\hline
\multicolumn{3}{@{}l}{{\bf X-ray}} \\ 
DQ Her variable type / low mass X-ray binary & P.6.1.8  (DQ, LMXB) & 30 & -&- & AOVMHW*\\
Close-binary super-soft source & P.6.2 (CBSS) & 1 &- &-& LS* \\
\hline
\multicolumn{3}{@{}l}{{\bf Other}} \\ 
Variable & P.7 & 1433  & -&- & LS\\
\hline
\end{tabular}
\end{table*}

\section{Algorithms}

Period finding algorithms can be divided into a number of types. The most popular seek to model a light curve via a least-squares fit to some set of (orthogonal) basis functions, most commonly trigonometric, such as Lomb-Scargle \cite{lomb, scargle} and its derivatives/extensions (e.g., \cite{gls}), though more complicated function sets, such as wavelets (\cite{foster}), have also been tried. Another approach is to minimize some measure of the dispersion of time series data in phase space, such as binned means (\cite{stellingwerf}), variance (\cite{aov89}), total distance between points (\cite{dworetsky}) or entropy (\cite{cnm95}), which can often be regarded as an expansion in terms of periodic orthogonal step functions. Bayesian methods (\cite{gregory92}, \cite{wang12}) are also becoming common and  
there have even been attempts to search for periodicity using neural networks (\cite{baluev}).

The basis of an algorithm also often determines how well it copes with the real world aspects of time series data, such as irregular sampling, gaps, and errors, e.g., standard Fourier analysis is impossible for any data diverging from regular sampling. \cite{dejager} argue that in the case of weak signals, most period finding methods only work well with certain kinds of periodic shapes and that this causes a selection effect for the general identification of weak periodic signals. Similar shape dependencies are found in \cite{schwarzenberg99}.

For this analysis, we have selected a representative set of the most common algorithms used which claim to be fast and accurate (see Table~\ref{methods} for operational details). This is a necessary condition for automated large-scale analyses of time series -- we consider an algorithm to be fast if it returns an answer in less than 5 s (assuming $\sim$250 points for a light curve and a $\sim$2 GHz CPU -- see section~\ref{performance for further discussion)}. This is a fairly conservative definition but roughly equates to analyzing 1000 light curves in just under 1.5 hours on a single processor. There are methods which can attain very high degrees of accuracy but do so at the expense of taking up to several minutes to work on a single time series, e.g., by using a very fine grain resolution in searching frequency space or involving a multistage process,  such as SuperSmoother (\cite{supersmoother}), \cite{jetsu} or \cite{shinbyun04}. These are well-suited for small-scale detailed analyses but not for the bulk processing that the new synoptic sky surveys warrant. However, for the sake of comparison, we have included SuperSmoother results for the MACHO data set, since it is largely regarded as the most accurate period finding technique.

Intuitively the fastest period finding algorithm will involve a single pass through a data set per trial period and integer counting operations, e.g., histogram binning. Any higher-order function calls, particularly per data point in a time series, will extend the average calculation time per trial period and, consequently, the overall time taken by the algorithm to determine a correct period. Of the algorithms considered, CE, AOV and PDM come closest to this ideal with a basic implementation employing integer arithmetic. AOV then requires two passes through a data set per trial period - one to compute the mean/variance of each bin, $\bar{x}_{i}$, and one to subtract the appropriate mean value from each data point, $x_{ij} - \bar{x}_i$. PDM is similar but CE only requires one pass.

One particular issue for automated period finders (particularly LS) is that they misidentify a multiple (or submultiple) of the period as the ``true'' period, i.e., the identified period, $p_i = m p_0$, where $m$ is an integer $n$ or its reciprocal, $1/n$, and $p_0$ is the correct period. This is a common problem for binary systems where the half period is frequently the most significant peak in a periodogram. For example, \cite{macc12} initially find 70\% of their periods for eclipsing binaries (EBs; $\sim$49\% of all objects) in the ASAS Catalog of Variable Stars (ACVS; \cite{acvs})  to be half periods. As discussed in \cite{wang12}, this is attributable to two aspects: for symmetric EBs, the true period and half its value are not clearly distinguishable quantitatively. Meanwhile, methods that are successful for EBs tend to find integer multiple periods of ``single bump'' stellar types, such as RR Lyrae and Cepheids, and vice versa. EBs also have two minima per cycle, while only one is expected by methods looking for sinusoidal-like variations. Clearly using a period (sub)harmonic instead of the true value can be a  problem for period-based statistics, such as Fourier decomposition where particular components would be assigned the wrong weights (amplitudes). We will consider the issue of period harmonics  further in section~\ref{results}.

\begin{table*}
\centering
\caption{Details of the various period finding algorithms used in this analysis. Where possible, we have used provided code, e.g., AOV/AOVMHW, PDM2, FastChi, and default parameter settings. The asterisk denotes those algorithms which were only applied to the MACHO data set.}
\label{methods}
\begin{tabular}{@{}llll}
\hline
Algorithm & Implementation & Behaviour & Reference\\
\hline
Lomb-Scargle (LS) & OpenCL$^{\dagger}$ & ${\cal O}(n^2)$ &  Lomb (1976); Scargle (1981); \\
& & & Townsend (2010) \\
Generalized Lomb-Scargle (GLS) & OpenCL$^{\dagger}$ & ${\cal O}(n^2)$ &  Zechmeister \& Kurster (2009) \\
Binned analysis of variance (AOV)$^{a,b}$ & F95 & ${\cal O}(nN)$ &   Schwarzenberg-Czerny (1989) \\
Multiharmonic analysis of variance  (AOVMHW)$^{a,c}$ & F95 & ${\cal O}(nN)$ & Schwarzenberg-Czerny (1996) \\
Phase dispersion minimization (PDM)$^d$ & F90 & ${\cal O}(nN)$  & Stellingwerf (1978) \\
Phase dispersion minimization (PDM2)$^{e,f}$ & C& ${\cal O}(nN)$ &  Stellingwerf (2011) \\
FastChi (FC)$^{g,h}$ & C & ${\cal O}(N \log N)$ &  Palmer (2009) \\
String length (STR) & F90 & ${\cal O}(nN)$ & Dworetsky (1983) \\
Conditional entropy (CE)$^i$ & F90 & ${\cal O}(nN)$ &  Graham et al. (2013)\\
Supersmoother (SS)$^{\ast j}$ & C& ${\cal O}(nN)$ &  Reimann (1994) \\
Correntropy kernel periodogram  (CKP)$^{\ast k}$ & C& ${\cal O}(n^{2})$ & Huijse et al. (2012) \\
\hline
\end{tabular}

\medskip
$^{\dagger}$ see Appendix~A; (a) http://users.camk.edu.pl/alex/soft/aovdist.tgz; (b) Overlapping phase bins; (c) 5 harmonics; (d) 10 phase bins; 

(e) http://www.stellingwerf.com/rfs-bin/index.cgi?action=PageView\&id=34; (f) Stellingwerf's improved algorithm; 

(g) http://public.lanl.gov/palmer/fastchi.html; (h) harmonics = 3, oversampling = 4; (i) 10 overlapping phase bins, 5 magnitude bins; 

(j) Code obtained from Andy Becker; (k) Code from Pablo Huijse
\end{table*}

\subsection{Frequency sampling}
\label{freqsamp}

The frequency sampling strategy used with a period finding algorithm is important. \cite{vio13} show that irregular sampling reduces the width of the peak at the correct frequency in the LS periodogram of a light curve if its temporal baseline is large. This means that there is a concrete risk of missing the peak if the periodogram is not computed for a sufficient large number of test frequencies. However, it also means that the error on the computed period will be less since this is also dependent on the width of the associated peak in the periodogram.

For a regularly sampled time series with time spacing, $\Delta t$, the Nyquist frequency, $\nu_{N} = 1 / 2 \Delta t$, constitutes an upper limit to the frequency range over which a periodogram can be uniquely calculated. For irregularly sampled time series, however, this value can be much higher - \cite{koen06} gives an upper limit of $0.5 \Delta$ for this frequency, where $\Delta$ is the best accuracy with which time is recorded. For example, in CRTS time is recorded to five decimal places giving $\nu_N = 5 \times 10^{-4} \mathrm{d}^{-1}$; in practice, though, a more manageable lower value would be used.

Two common frequency gridding strategies are applied in the literature when working with large collections of time series. The first (\cite{macc12}, \cite{deboss07}) uses for all light curves: $\nu_{\min} = 0$, $\nu_{\max} = 10$, and $\delta\nu = 0.1/\Delta\tau$, where $\Delta\tau$ is the data timespan. The second (\cite{aovmhw}) estimates optimal values for each light curve: $\nu_{\min} = 0$ or $\delta\nu$, $\nu_{\max} = 1 / 2 \tau_{\mathrm med}$, and $\delta\nu = 1 / (A \times \Delta\tau)$, where $\tau_{\mathrm med}$ is the median difference between successive ordered times, $A$ is a factor, typically 10 - 15,  taking into account oversampling and binning or the number of harmonics used in a Fourier fit, and $\Delta\tau$ is the data timespan.

We have applied both the optimal strategy and fixed $\delta \nu$ values of $\delta \nu = 0.0001$, $0.001$, and $0.01$ over a frequency range with $\nu_{\min} = 0$ and $\nu_{\max} = 20$. The median timespan for all the data is $\sim2618$ days which gives a median $\delta\nu = 2.5 \times 10^{-5} \mathrm{d}^{-1}$, assuming $A=15$ in the optimal case. We also note that many algorithms use a finer resolution grid to get a more accurate period estimate once a primary peak has been found with the coarse grid.

\section{Metrics}
For the purposes of this analysis, we define three metrics: one to evaluate the accuracy of the recovered period of a light curve compared to the value of its true period and two to measure the quality of a light curve.

\subsection{Accuracy metric}
\cite{lsstrrl} define a matching criterion for period recovery using the quality of the period-folded data as a metric:

\[ \frac{|P_{al} - P_{in}|}{P_{in}} \le \frac{\delta \phi_{max} P_{in}} {\Delta \tau} \]

\noindent
where $P_{in}$ is the known input period, $P_{al}$ is the period according the algorithm under investigation, $\Delta \tau$ is the duration of the time series and $\delta \phi_{max}$ is the maximum allowed phase offset after period-folding $N$ cycles. For simulated RRab light curves in LSST, this translates to:

\[ \frac{|P_{al} - P_{in}|}{P_{in}^2} \le  10^{-5} \mathrm{day}^{-1} \]

\noindent
for a maximum period-folded phase offset of 1/27th of a cycle or a period within $\sim0.22s$ of the true value for a 0.5d period star and a 10 year survey. Given the variation in the baselines of the light curves in this analysis, particularly within the CRTS data set where there is a dependency on both galactic latitude and which telescope was used to observe the object, a fixed survey length makes little sense. Instead for each object, we consider its temporal coverage but keep $\delta \phi_{max} = 1/27$ so that a 10-year baseline will give equivalent accuracy to LSST. We will use the equivalent accuracy level of $10^{-5}$ day$^{-1}$ as a fiducial value. For the median baselines of the three surveys, the corresponding accuracy values are $1.4 \times 10^{-5} \mathrm{day}^{-1}$ for ASAS and MACHO and  $1.7 \times 10^{-5} \mathrm{day}^{-1}$ for CRTS. 

\cite{dubath} consider a period as good if the difference between the calculated period (using LS/GLS) and the quoted value leads to a cumulative shift in phase of less than 20\% over the full timespan of the light curve. This equates to an accuracy level of $\sim 10^{-4} \mathrm {day}^{-1}$ for a 10 year baseline. Meanwhile, \cite{macc12} claimed 77.2\% exact agreement between the periods they found for ASAS objects (using a generalized LS-based algorithm) and those given by ACVS. However, in terms of the matching criterion used here, only 20.2\% of the periods actually agree between the two sets at the $10^{-5} \mathrm{day}^{-1}$ accuracy level. The quoted agreement of Richards et al. is found at an accuracy level of $\sim10^{-3} \mathrm {day}^{-1}$.

We have therefore considered equivalent accuracy cutoffs for a 10-year baseline of $10^{-3}$ ($\delta\phi_{max} = 100/27$), $10^{-4}$ ($\delta\phi_{max} = 10/27)$ and $10^{-5}  \mathrm {day}^{-1}$ ($\delta\phi_{max} = 1/27)$ respectively for our comparisons to reflect the range used in the literature and, for the value of $10^{-5}$, with a view to future surveys. We note that the error in the determined period could be larger than a particular accuracy cutoff. However, as already noted, most of the algorithms in this analysis use a finer grain resolution to get a more accurate estimate once an initial value has been found with a coarser grain resolution. This is typically a factor of a hundred smaller than the coarse grain resolution step and in this analysis would be a maximum of $\Delta \nu = 10^{-5}$. The effect of this should therefore be minimal with reference to a particular cutoff value.

We also want to have an accuracy metric relevant for period harmonics since periodicity in an object can still be detected, even if only a harmonic of the true period is found (\cite{ckp}). We modify the criteria used by \cite{ckp} so that an accurate harmonic is identified if:

\[ \left| \frac{P_{al}}{P_{in}} - \left\|\frac{P_{al}}{P_{in}}\right\| \, \right| < \frac{\delta \phi_{max} P_{in}} {\Delta \tau}  \; \mathrm{for} \; P_{al} > P_{in} \]

\noindent
and:

\[ \left| \frac{P_{in}}{P_{al}} - \left\|\frac{P_{in}}{P_{al}}\right\| \, \right| < \frac{\delta \phi_{max} P_{in}} {\Delta \tau}  \; \mathrm{for} \; P_{al} < P_{in} \]

\noindent
where $\|x\|$ is the nearest integer to $x$ relative to the same accuracy cutoff used for periods.

\subsection{Quality measure}
There are several sets of light curves of the same class of variable object that we would like to compare on a common quality basis but they have been produced by different telescopes and so span different magnitude ranges; for example, an RR Lyrae light curve in the ASAS data set has the same subjective quality, i.e., visually appears the same in terms of error size and scatter, at 12$^{\mathrm{th}}$ magnitude as a 20$^{\mathrm{th}}$ magnitude light curve in the MACHO data set. 

\begin{figure}
\caption{This shows the overall distribution of the signal-to-noise ratio for all the light curves and the stacked relative contributions of each of the individual data sets: ASAS (red), CRTS (blue) and MACHO (black).}
\label{snrhist}
\includegraphics[width=3.4in]{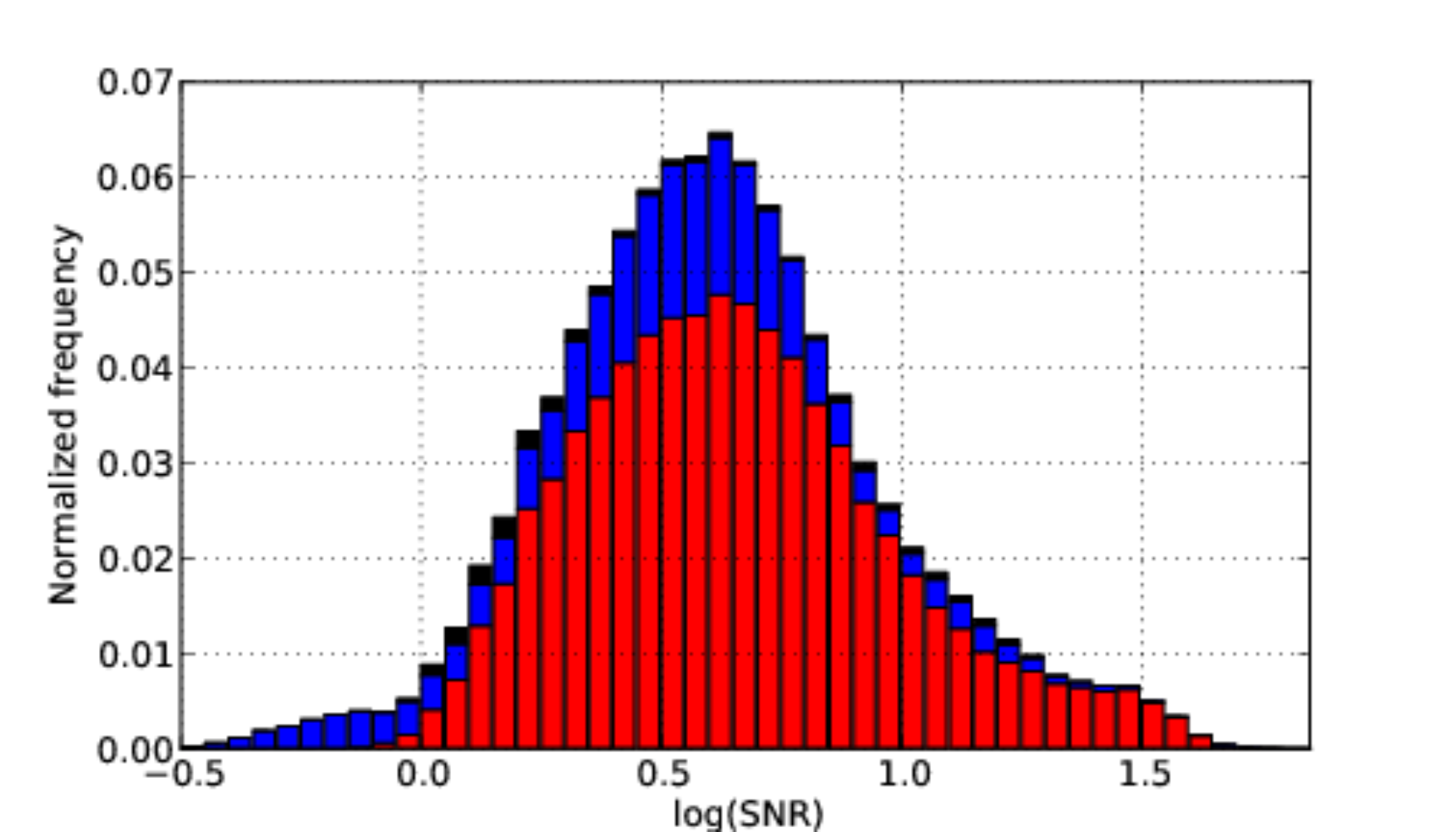}
\end{figure}

The signal-to-noise ratio (SNR; mean of signal / standard deviation of noise) provides a general matching criterion. \cite{rimoldini} gives an expression for the SNR of a light curve and we employ a slightly modified form here:

\[
\mathrm{SNR} = \left[ \frac{\sum_{i=1}^{n}w_i(x_i - x_m)^2 + \sum_{i=1}^{n}w_i^2 \epsilon_i^2 / W}{\sum_{i=1}^n w_i \epsilon_i^2} \right]^{1/2}
\]

\noindent 
where $x_m$ is the median magnitude (instead of the mean), $\epsilon_i$ the photometric error of the $i_{th}$ data value and $W = \sum_i w_i$. We employ $w_i = 1$ in this analysis. Fig~\ref{snrhist} shows the distribution of SNR values for the three surveys considered here. 

We note, though, that this measure is based on mean quantities and that changes in the overall shape of a light curve for different object types will have an impact: for example, there is a strong correlation between the SNR of a light curve and the amplitude of its variability and this is also survey dependent (see Fig.~\ref{snrmad}). Since SNR is essentially a measure of the intrinsic scatter within a data set, it is conceptually similar to the entropy. Standard estimators for entropy, though, are optimized for signal detection and do not take into account the contributions of noise. \cite{cincotta99} defines a modified estimator for the Shannon entropy of a data set which takes observational errors into account and we will use this as class-based quality comparisons. The Shannon entropy, $H_0$, for a distribution on the unit square partitioned into $k$ partitions is:

\begin{figure}
\caption{This shows the distributions of the signal-to-noise ratio vs. the median absolute deviation (from the median) for the three surveys: ASAS (red), CRTS (blue) and MACHO (black). There is clearly a strong correlation between SNR and the amplitude of variability and this is a survey dependent effect - the same SNR value equates to a different range of variabilty for each survey.}
\label{snrmad}
\includegraphics[width=3.4in]{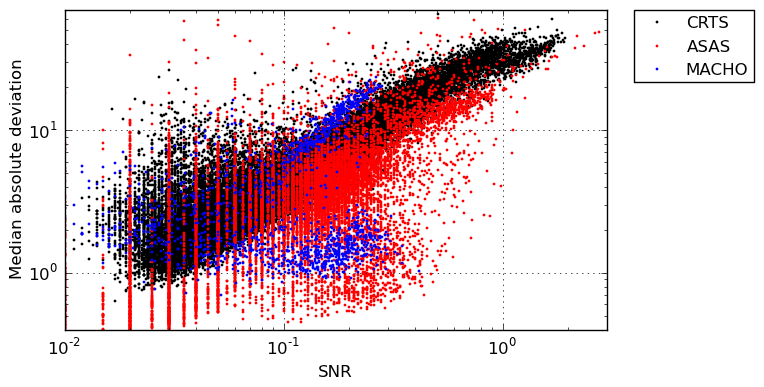}
\end{figure}

\[
H_0 = - \sum^{k}_{i = 1} \mu_i \ln(\mu_i); \forall \mu_i \ne 0
\]

\noindent
where $\mu_i$ is the occupation probability for the $i^{th}$ partition. For a data set where measurement $v_i$ has an error $\epsilon_i$, the occupation probability is given by:

\[
\widetilde{\mu} = \frac{R}{2} \sum^{n_l}_{i=1} [ \mathrm{erf}(w_{im} + \Delta w_i) - \mathrm{erf}(w_{im}) ]
\]

\noindent
where $\mathrm{erf}(x)$ is the error function, $n_l \simeq N/L$ is the number of points in the $l^{th}$ partition of the $x$-axis (i.e. with phase in the $[l/L, (l+1)/L]$ interval), and

\[
w_{im} = \frac{m - Mv_i}{\sqrt{2}M\epsilon_i}, \Delta w_i = \frac{1}{\sqrt{2}M\epsilon_i}, 
\]\[
\frac{2}{R} = \sum_{i=1}^{N}[\mathrm{erf}(w_{iM}) + \mathrm{erf}(| w_{i0} |)]
\]

\noindent
and $L$ and $M$ are the number of partitions along the $x$-axis and $y$-axis respectively. 

\begin{figure}
\caption{This shows the distribution of the median photometric error and the modified entropy for all the light curves phased at their quoted period considered in this paper. The three surveys are denoted by red (CRTS), blue (MACHO) and black (ASAS). The artifact at mean error = 0.05 in the CRTS data set results from a lower limit to error size in this data set. The small median errors at the highest entropy level may indicate non-monoperiodic (multiperiodic, irregular, etc.) sources.}
\label{mederr_ent}
\includegraphics[width=3.34in]{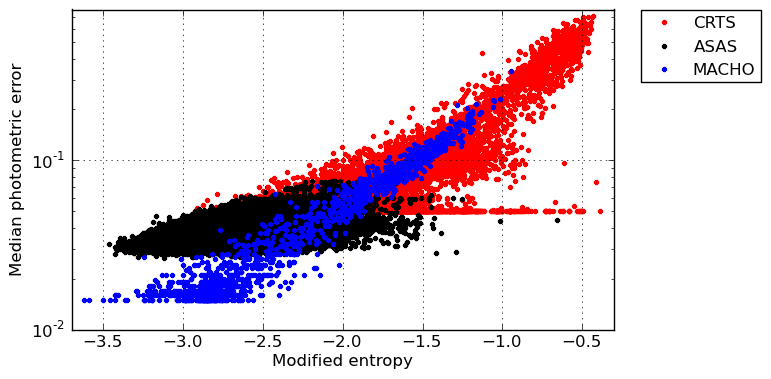}
\end{figure}

Fig~\ref{mederr_ent} shows the distribution of the modified entropy for all the light curves, phased at their quoted periods, in the CRTS, ASAS and MACHO data sets against their median photometric errors. This shows that there is a general relationship between the two quantities and therefore light curves with the same modified entropy can be considered to be qualitatively similar (in broad signal-to-noise terms) and therefore compared on an equal footing. Note that the computational cost of the estimator - here involving error function calculations - is too high for consideration as a viable period finding algorithm in this paper.

\section{Results and analysis}
\label{results}

The efficacy of the different period finding algorithms is clearly dependent on a number of factors.
We evaluated each of the algorithms in terms of completeness, i.e., the fraction of true periods recovered within a defined accuracy limit, as a function of various quantities. The two most obvious variables to consider are magnitude and the number of observations in a light curve. The resolution used to scan the range of trial periods (frequencies) will also have an effect. The variability of a light curve, both naturally due to the actual variability of the source and acquired as measurement scatter (noise), as well as the actual class of variable object (and the shape of the light curve) and its period are also factors to evaluate. Finally, the actual time taken to determine an accurate period can be an important aspect in determining the {\em usability} of particular algorithms in addition to their accuracy.

\subsection{Magnitude}
\begin{figure*}
\centering
\caption{This shows the completeness fraction for the different period finding algorithms as a function of magnitude for each data set: (a) MACHO, (b) CRTS, and (c) ASAS. The three plots in each row are for different accuracy cutoffs, equivalent from the left to 10$^{-5}$, 10$^{-4}$ and 10$^{-3} \, \mathrm{days}^{-1}$ over a 10-year baseline respectively (see text). The different algorithms are denoted by: AOVMHW (blue diamonds), AOV (green diamonds), CE (black circles), LS (green triangles), PDM (left-facing blue triangles), PDM2 (cyan inverted triangles), GLS (right-facing orange triangles), FC (magenta squares), SS (red stars), CKP (yellow stars), and STR (yellow pentagons). The optimal frequency sampling was used where relevant. The small red dots indicate the cumulative magnitude distribution of the relevant data set.}
\label{res_mag}
\includegraphics[width=7.0in]{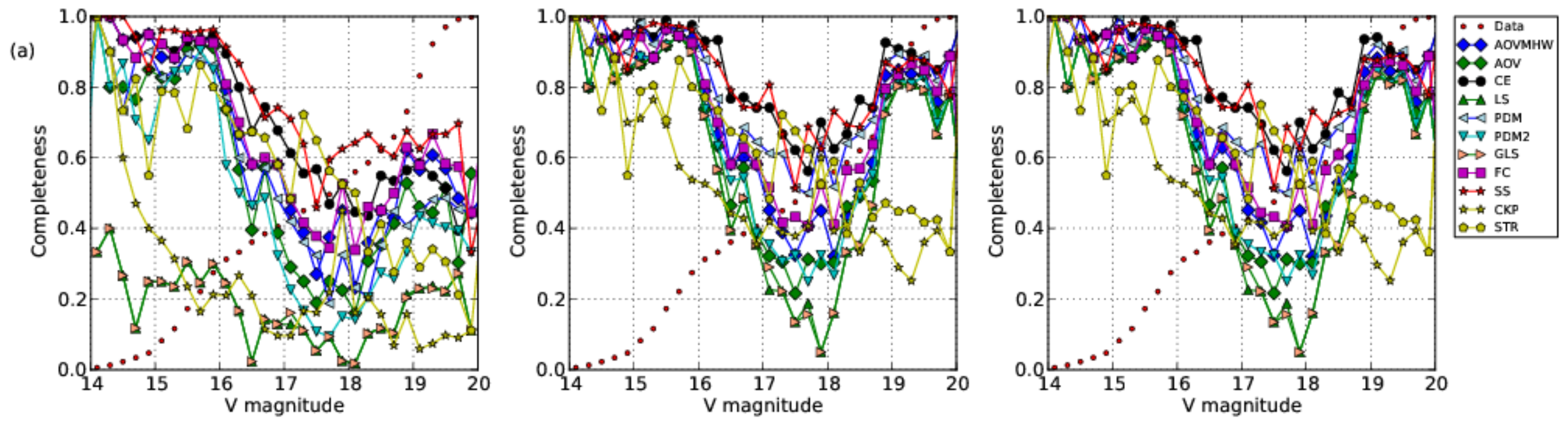}
\includegraphics[width=7.0in]{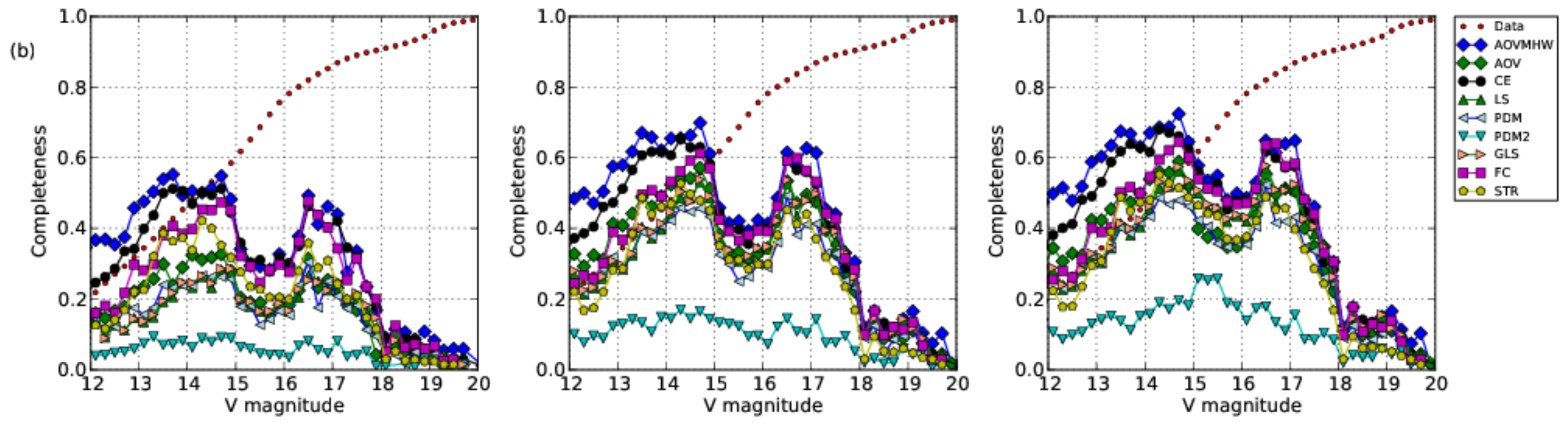}
\includegraphics[width=7.0in]{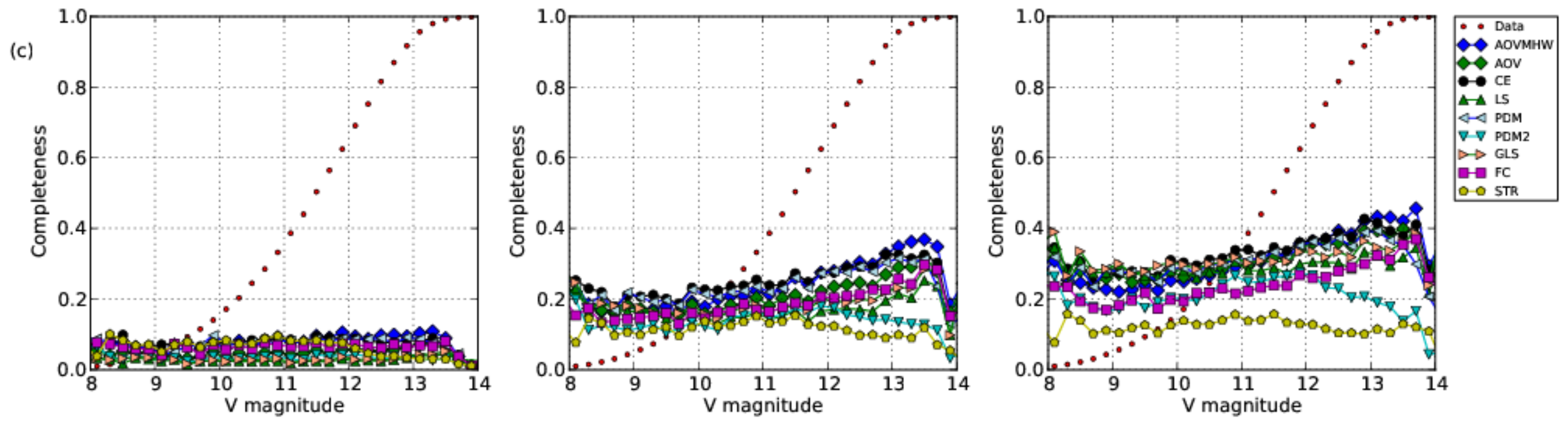}
\end{figure*}

Fig.~\ref{res_mag} shows the completeness fraction as a function of magnitude for the three data sets and accuracy cutoffs respectively. With the MACHO and CRTS data, there is a general decline in accuracy with increasing magnitude, particularly past the 90th percentile in the magnitude distribution, as the photometric errors become more significant and the light curves noisier. There are also dips in both data sets around the 60th percentile which is most likely connected to the relative magnitude distributions of different classes of object; for example, in the MACHO data, $\sim \! 90$\% of the objects between 14th and 16th magnitude are Cepheids, whereas between 16th and 18th magnitude, there are twice as many eclipsing binaries as Cepheids. The ASAS data seems fairly flat except at its very faintest end. With this data, 
magnitude is weakly correlated with SNR, i.e., fainter objects at fainter magnitudes tend to have slightly 
better quality light curves and the methods are more likely to recover the true period for an object with a 
higher SNR (see section~\ref{quality}). The combination of the two in this case gives a fairly constant relationship between completeness and magnitude. The low levels of completeness relative to the other two data sets are a consequence of the large number of semi-regular variables and similar pulsating objects in this data set ($\sim \! 50$\%) for which accurate periods could not be established (see below). 

The comparatively better performance of AOVMHW and CE at brighter magnitudes with the CRTS data set indicates that these algorithms work well with data which may contain saturated values. The nominal saturation limit for CRTS is $V \sim 12$ and the magnitude used in this analysis is the mean magnitude of the light curve so there may well be observations of bright objects near maxima which are saturated. These algorithms are not attempting to model the phased light curve as a sum of sinusoidal functions and so are less susceptible to nonsinusoidal or truncated sinusoidal-shaped light curves that may occur near a survey's saturation limit.
The poor performance of PDM2 with this data set indicates an issue with the irregular sampling strategy of CRTS light curves relative to those in the other two data sets.

\subsection{Observations}
\begin{figure*}
\centering
\caption{This shows the completeness fraction for the different period finding algorithms as a function of the number of observations per time series for each data set: (a) MACHO, (b) CRTS and (c) ASAS. The three plots in each row again correspond to the different accuracy cutoffs: $10^{-5}$, $10^{-4}$, and $10^{-3} \, \mathrm{days}^{-1}$ over a 10-year baseline. The same symbols are used for each algorithm as in Fig.~\ref{res_mag} with the optimal frequency sampling used where relevant. }
\label{res_obs}
\includegraphics[width=7.1in]{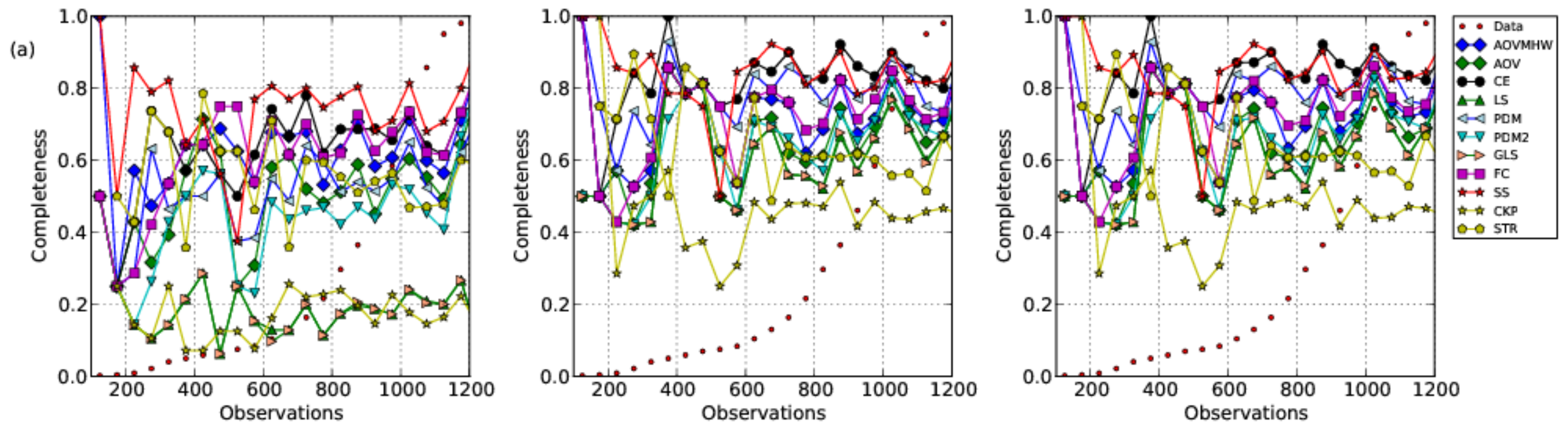}
\includegraphics[width=7.1in]{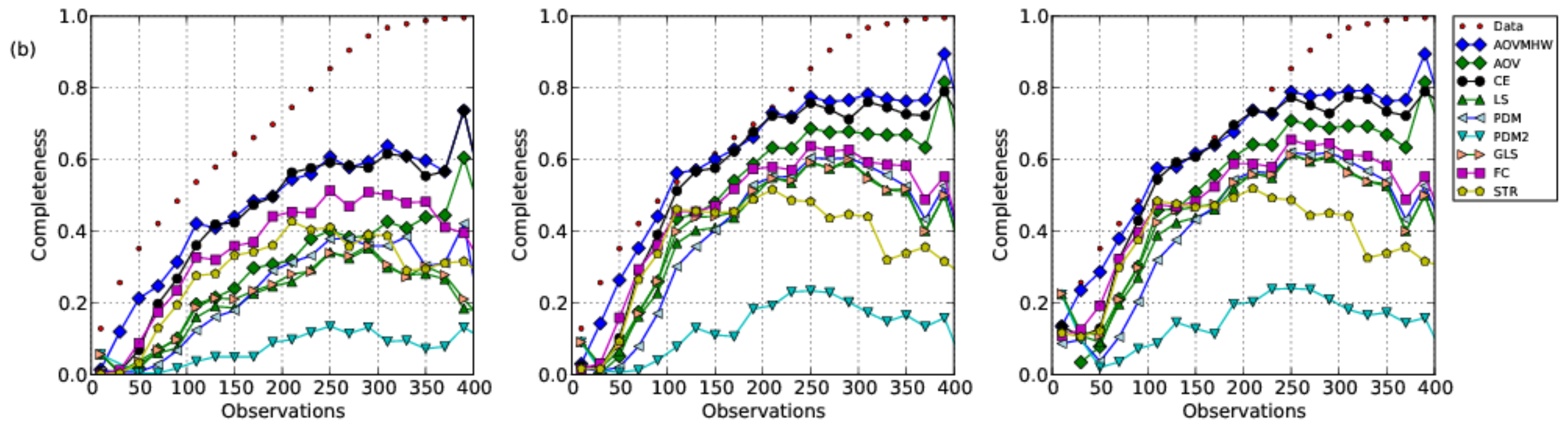}
\includegraphics[width=7.1in]{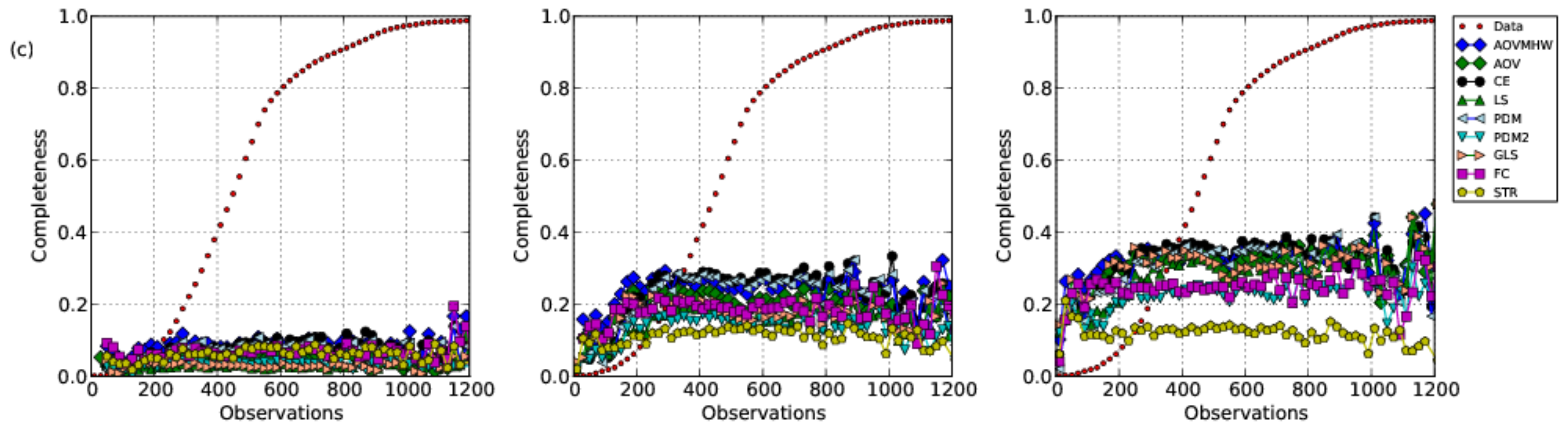}
\end{figure*}

Fig.~\ref{res_obs} shows the completeness fraction as a function of number of observations, $n$, for the three data sets and accuracy cutoffs. The MACHO and ASAS data sets show no strong dependency on the number of observations, except possibly at smaller values ($n < 200)$. However, the CRTS data show a definite dependency. For $n < 200$, there is generally insufficient coverage or sampling of phase for the algorithms to detect the true period effectively. This is compounded by the observing strategy of the CRTS survey: a set of 4 observations, each separated by 10 minutes, repeated once or twice per lunation. 

\begin{figure}
\centering
\caption{This shows the distribution of time differences in days between successive observations for the three data sets: ASAS (red), MACHO (black) and CRTS (blue). The bin widths are 0.1 dex in log(t).}
\label{dt}
\includegraphics[width=3.4in]{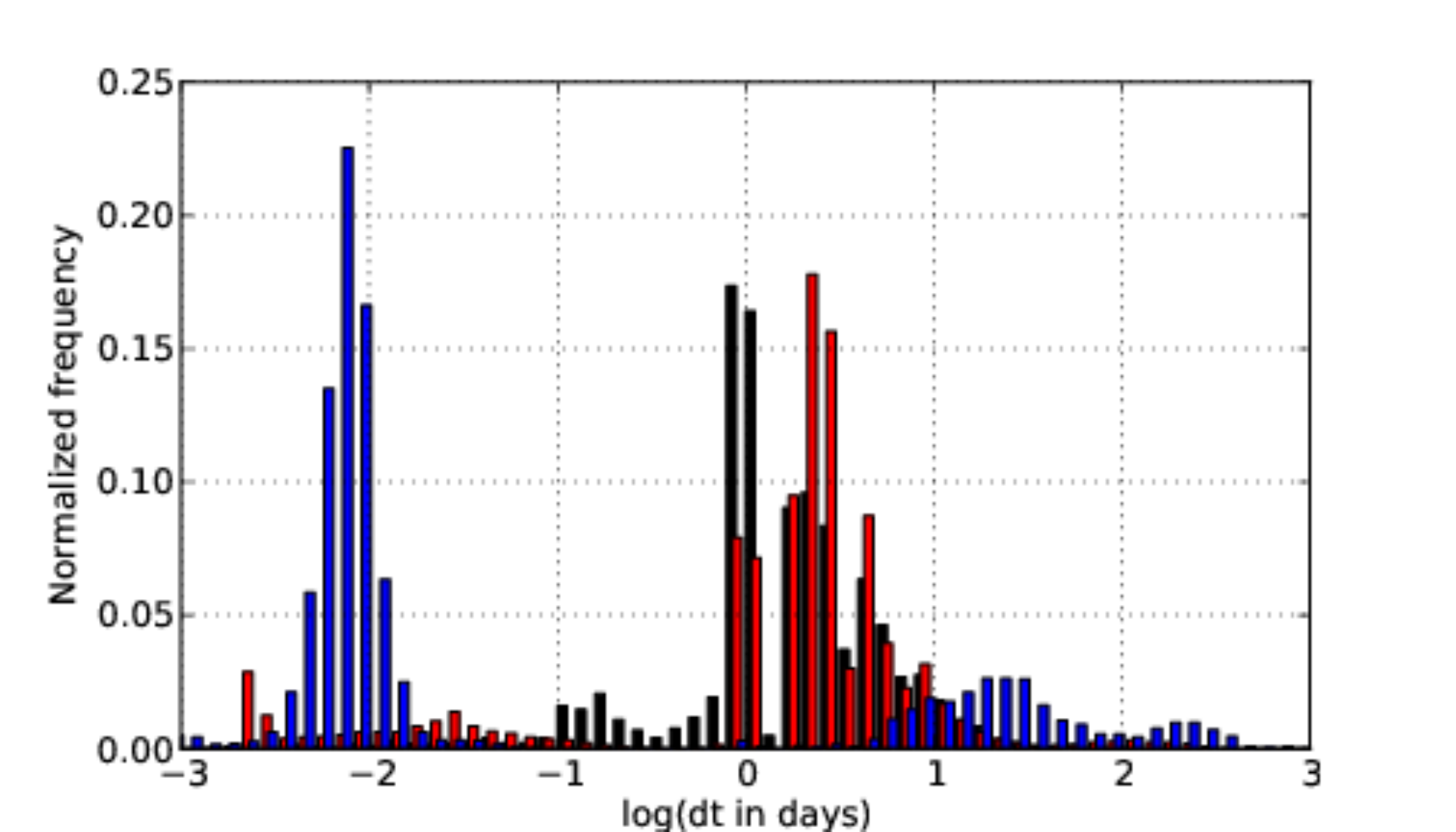}
\end{figure}

Fig.~\ref{dt} shows the distribution of the time difference, $dt$, in days between successive observations for the three data sets considered here. From these distributions, we can estimate the number of observations that would be required to ensure a particular minimum phase coverage density for an object of a given periodicity. For each data set, we generate a random observing schedule (set of successive observations) drawn from the appropriate distribution and then determine what the corresponding phased light curve coverage would be for a particular test period in terms of the minimum bin occupancy assuming bin widths of $\Delta \phi = 0.1$.
Table~\ref{optobs} gives the median number of observations per time series from 5000 simulations of each data set over a frequency ($1. / $ period) range of 0 - 4 and for minimum bin occupancies of 5, 10 and 20 observations respectively.

\begin{table}
\center
\caption{The median number of observations required to ensure the specified minimum coverage in the binned phased light curve of an object for each data set, assuming bin widths of $\Delta \phi = 0.1$}
\label{optobs}
\begin{tabular}{lccc}
\hline
Sampling distribution & \multicolumn{3}{|c|}{Minimum bin occupancy} \\
 &  5 & 10 & 20 \\ 
\hline
ASAS & 90 & 155 & 276\\
MACHO & 87 & 150 & 270\\
CRTS & 138 & 214 & 350\\
\hline
\end{tabular}
\end{table}

The bimodal observing distribution of dt of the CRTS data set (see Fig.~\ref{dt}) means that a larger number of individual observations are required for the same phase coverage relative to the other distributions. However, this requires fewer actual nights since each night provides four individual observations. This observing strategy also provides greater sensitivity to short timescale phenomena - \cite{vio13} show that irregular sampling permits one to retrieve information about frequencies much greater than the Nyquist frequency. We note that the proposed core LSST observing strategy is very similar with two back-to-back 15~s exposures and a return to the same pointing within 15-60 minutes, giving four observations within an hour (\cite{lsstrrl}).

It may still be the case, however, that there is not enough baseline in any of the surveys to accurately establish the periods of objects with very long periods. If we assume a minimum bin occupancy in phase space of $b$ per bin of width $\Delta \phi$ then {\em regular} sampling of an object with period $P$ would require an observation every $\Delta t = P \Delta \phi / b$ days. The total number of observations in a light curve, $n$, for a survey with baseline $\tau$ would then be given by $n = \tau / \Delta t$. Rearranging this gives the minimum baseline for a survey to adequately sample a light curve as: $\tau \geq P n \Delta \phi / b $. For an object with a period of 2000 days, say, which is observed regularly to ensure a minimum bin occupancy of 10 with $\Delta \phi = 0.1$ bin widths, the minimum baseline with 150 observations would be 3000 days. 

\begin{figure*}
\centering
\caption{This shows the completeness fraction for the different period finding algorithms as a function of the quoted period in days for each data set: (a) MACHO, (b) CRTS and (c) ASAS. The three plots in each row again correspond to the different accuracy cutoffs: $10^{-5}$, $10^{-4}$, and $10^{-3} \, \mathrm{days}^{-1}$ over a 10-year baseline. The same symbols are used for each algorithm as in Fig.~\ref{res_mag} with the optimal frequency sampling used where relevant.}
\label{per_obs}
\includegraphics[width=7.0in]{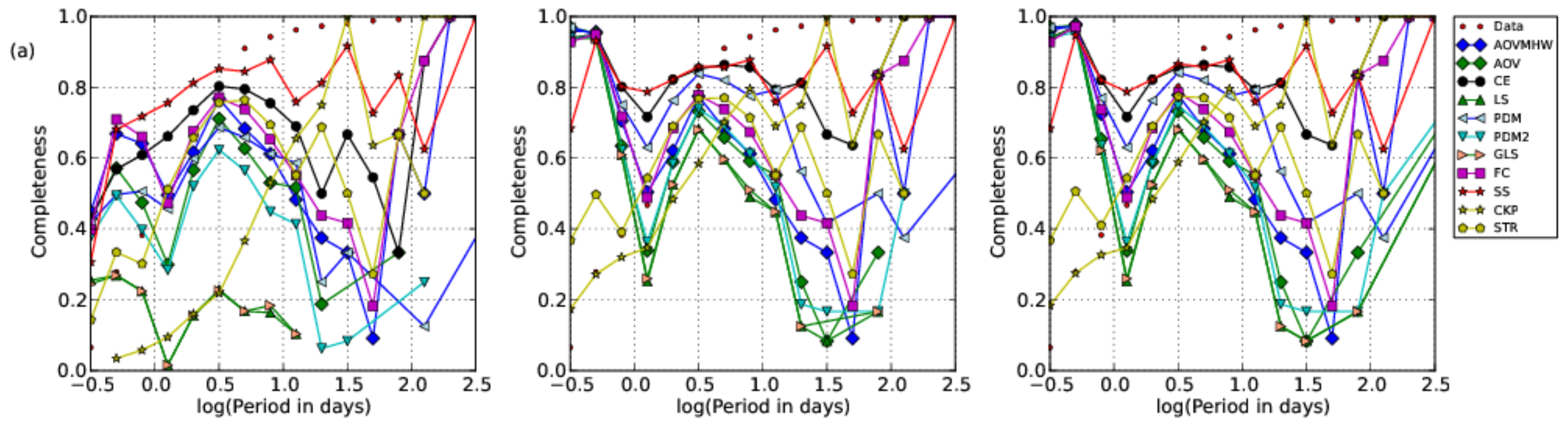}
\includegraphics[width=7.0in]{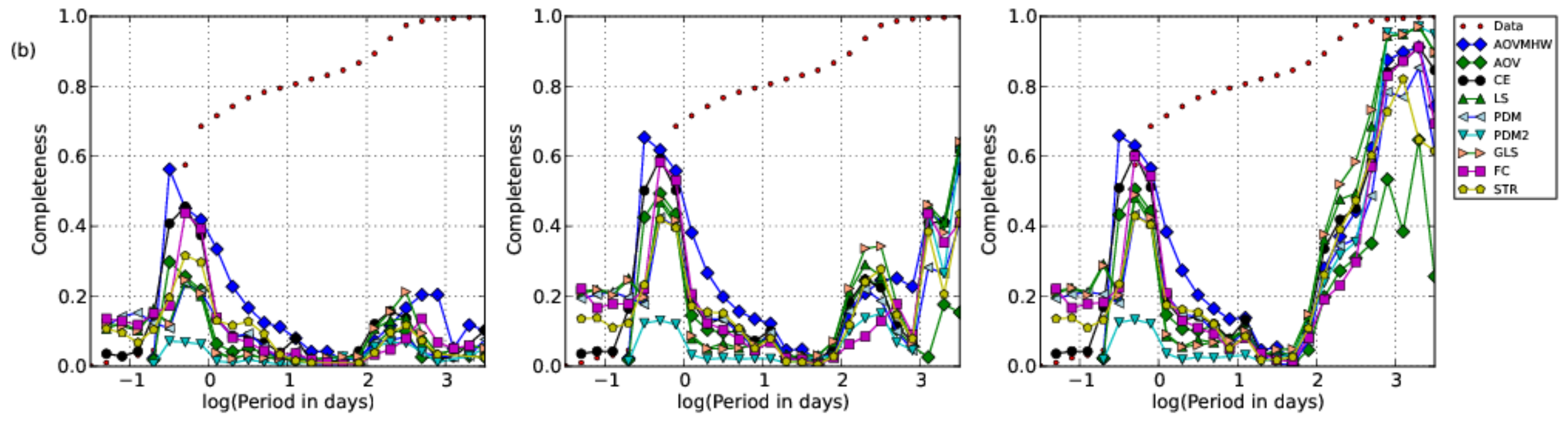}
\includegraphics[width=7.0in]{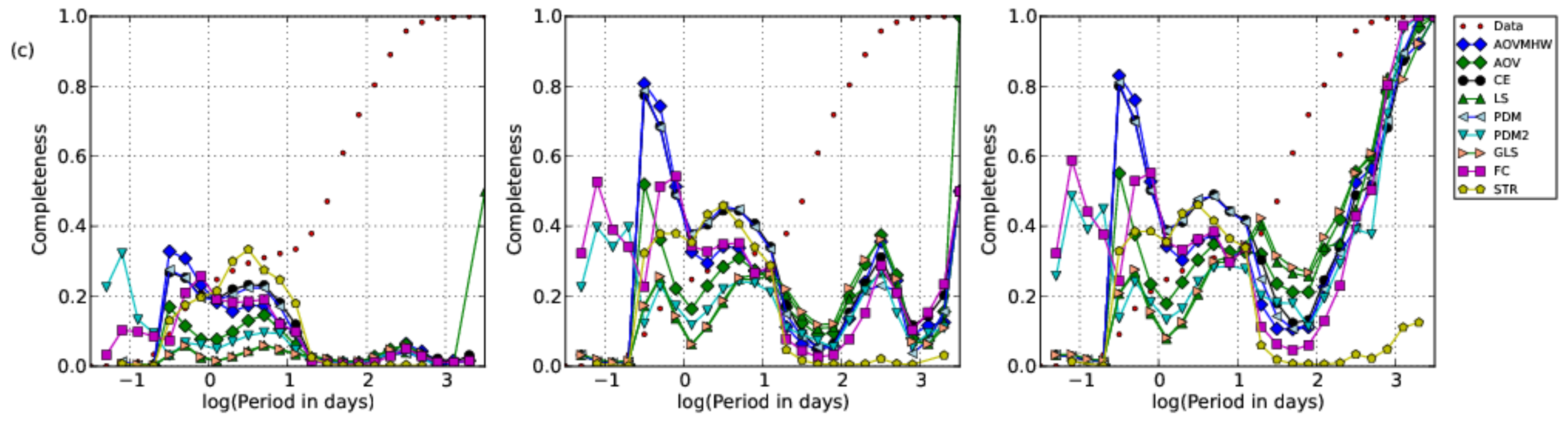}
\end{figure*}

Fig.~\ref{per_obs} shows the results as a function of the quoted period, $p$, for the three data sets and accuracy cutoffs. All three surveys have baselines $>2000$ days and this is clearly sufficient to recover periods with an accuracy cutoff of $10^{-3}$ for long period objects but less so if higher degrees of accuracy are required. The overall lack of performance for objects with period between roughly 10 and 100 days is due to the (in)efficiencies of the methods with the particular classes of object with those period lengths (see section 4.3.4).

In terms of both the number of observations and the quoted period, the relative performances of the period finding algorithms seen as a function of magnitude in the previous section are also repeated here with AOVMHW and CE the most successful. As well PDM2 again shows the same issues with CRTS data. We also infer that all the period finding algorithms are stable with a minimum bin occupancy of $\sim \!\! 10$, assuming bin widths of $\Delta \phi = 0.1$. 

\subsection{Resolution and quality}
\label{quality}
\begin{figure*}
\centering
\caption{This shows the completeness fraction for the combined data set for the seven algorithms where multiple frequency sampling strategies were applied: AOVMHW, AOV, CE, STR, GLS, LS, and PDM. The four curves per plot are: optimal $\delta \nu$ (blue diamonds), $\delta \nu = 0.0001$ (green triangles), $\delta \nu = 0.001$ (inverted cyan triangles), and $\delta \nu = 0.01$ (magenta squares) respectively. An accuracy cutoff of $10^{-3}$ was used for greatest contrast. The quality of the light curves improves from left to right, i.e., there is less acquired scatter in a light curve with increasing SNR. The red dots indicate the cumulative SNR distribution of the combined data set.}
\label{freqsampcomp}
\includegraphics[width=7.1in]{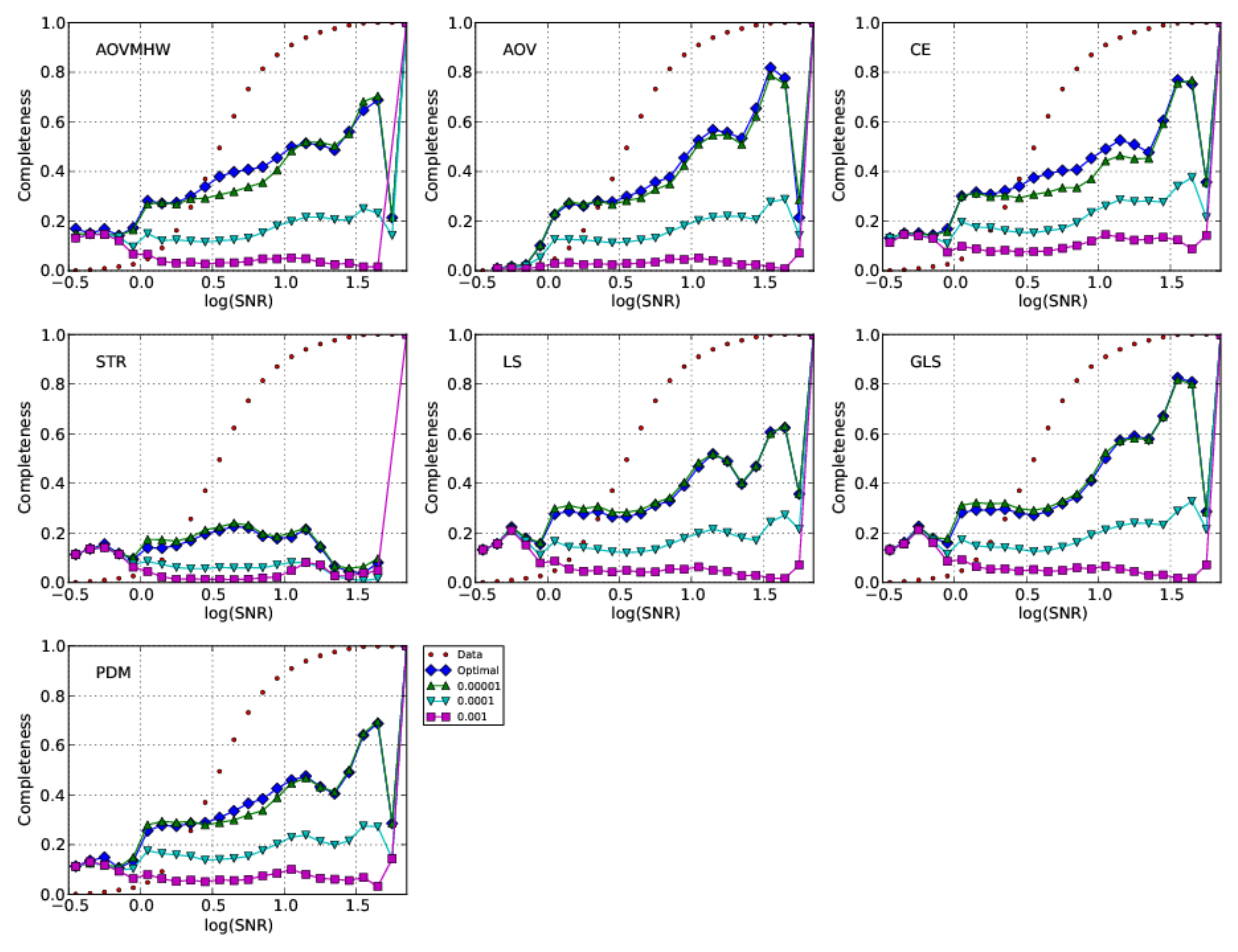}
\end{figure*}

We have combined the three data sets in terms of our quality measure (modified entropy). Fig.~\ref{freqsampcomp} shows the results for this data for each algorithm that allowed the frequency resolution to be set --  AOVMHW, AOV, CE, STR, LS, GLS, and PDM -- and the different frequency resolutions employed. 
At the accuracy cutoff used ($10^{-3}$ - note that any possible effects of errors in the derived period will be two magnitudes smaller than the cutoff value), there is very little difference between the performance of $\delta \nu = 0.0001$ and the optimal $\delta \nu$ (as noted in Section~ \ref{freqsamp}, the median optimal $\delta \nu$ is $2.5 \times 10^{-5}$) for all the algorithms  considered. This suggests that computation time can be saved in future by using a standard frequency resolution of $\delta \nu = 0.0001$ in (initial) frequency range scans and then a finer/optimal resolution for higher accuracy if required. If a lower resolution is preferred then the CE algorithm gives the best performance relative to the others, even for $\delta \nu = 0.01$. 

The overall performance of the algorithms as the quality of the light curves varies is broadly consistent. None of the algorithms work with the noisiest of light curves, i.e., those showing the most acquired scatter as opposed to variability. All methods show a peak at $\widetilde{\mu} \sim -2$ in the best resolution curves, corresponding to RR Lyrae stars, and a slight hump at $\widetilde{\mu} \sim 2.5$ from eclipsing variables. The best quality light curves ($\widetilde{\mu} < -3$) are dominated by semiregular and pulsating red giant variables. The slightly better relative performance of the best frequency resolution LS and GLS algorithms with these classes is most likely related to the quoted periods for these objects  also having been determined with these algorithms. We will discuss this more in the next subsection. 

\subsection{Class}
\begin{figure*}
\centering
\caption{This shows the completeness fraction for the different period finding algorithms on the full combined data set in terms of the seven different broadest classes of variable object represented: P.1 (eruptive), P.2 (pulsating), P.3 (rotating), P.4 (cataclysmic), P.5 (eclipsing), P.6 (X-ray), and P.7 (other). The same symbols are used for each algorithm as in Fig.~\ref{res_mag} with the optimal frequency sampling used where relevant. An accuracy cutoff of $10^{-4}$ was used.}
\label{clscomp}
\includegraphics[width=7.1in]{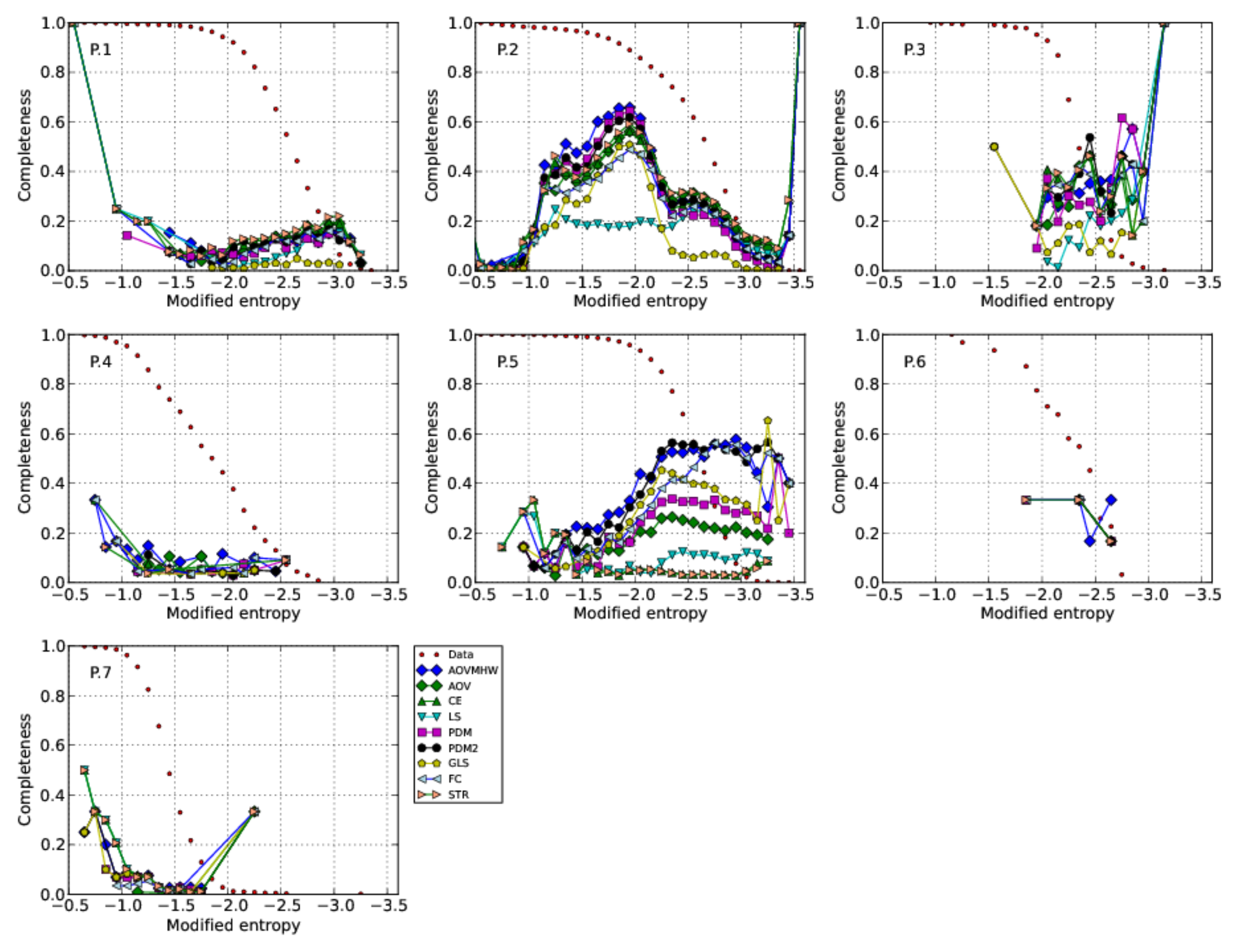}
\end{figure*}

The combined data set can also be considered in terms of the various classes of object represented in the data. Fig.~\ref{clscomp} shows the results as a function of modified entropy for each of the broadest class designations used: eruptive (P.1, 4194 objects), pulsating (P.2, 45599 objects), rotating (P.3, 455 objects), cataclysmic (P.4, 386 objects), eclipsing (P.5, 14952 objects), X-ray (P.6, 31 objects), and other (P.7, 1434 objects). Unsurprisingly the best results are obtained for the pulsating and eclipsing variable classes as these contain the best defined periodic objects; however, the periods of rotating objects can also be recovered to a reasonable degree. The poor performance for the other classes is most likely caused by a general lack of any clear periodic signal in the light curves for these types of object, for example, LPVs do not seem to oscillate in a clean fashion and so their periods are intrinsically not very well defined.

The shapes of these curves can be attributed to the relative contributions made by different subclasses of object within each of the broadest classes. For example, within the pulsating class, classical Cepheids (Delta Cep) have a mean $\widetilde{\mu} = -1.22$, RR Lyrae have $\widetilde{\mu} = -1.96$, Mira have $\widetilde{\mu} = -2.37$, and semiregular variables have $\widetilde{\mu} = -2.74$. Similarly, within the eclipsing class, there is a sequence from AM Her variables to Algol types to Beta Lyrae types to W UMa types, although the performance for the three eclipsing binary classes is fairly constant. The peak at $\widetilde{\mu} \sim 3$ in the eruptive class results corresponds to weak-line T Tauri stars.

\begin{figure*}
\centering
\caption{This shows the completeness fraction for the different period finding algorithms for eclipsing variables (P.5) in the full combined data set using just strict period matching (left plot) and allowing period (sub)harmonics as well (right plot). The same symbols are used for each algorithm as in Fig.~\ref{res_mag} with the optimal frequency sampling used where relevant. An accuracy cutoff of $10^{-4}$ was used.}
\label{p5comp}
\includegraphics[width=7.1in]{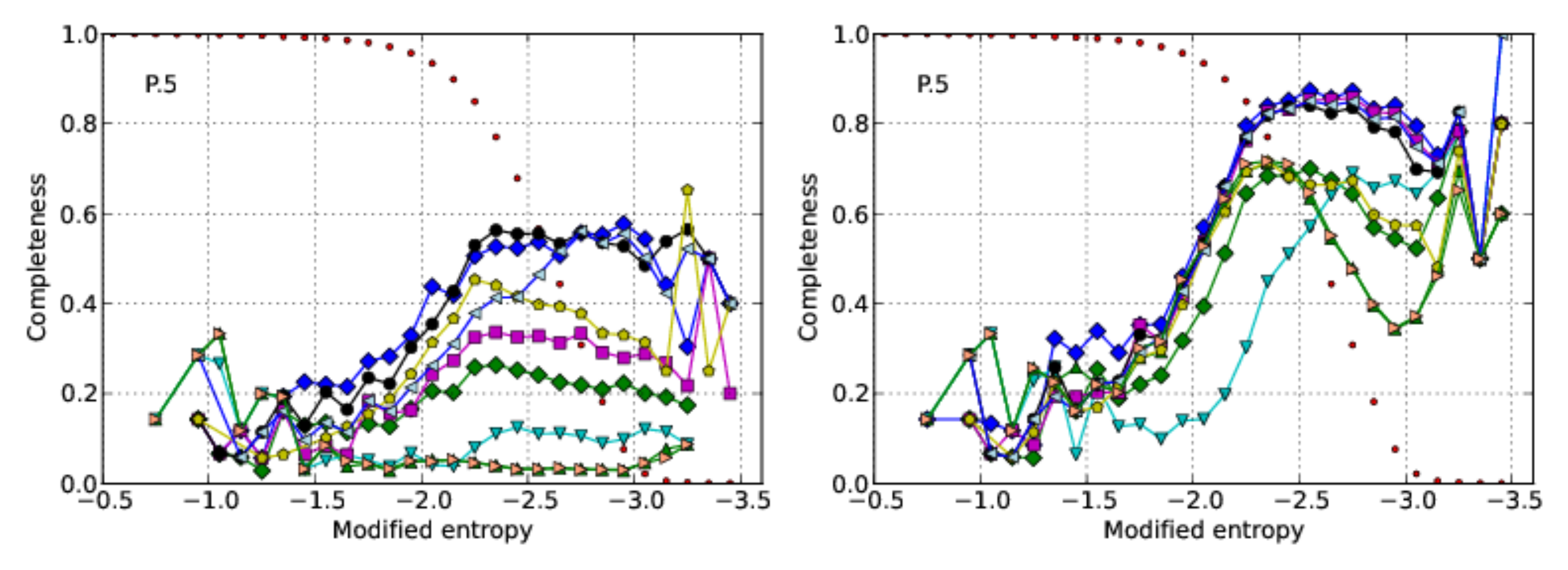}
\end{figure*}

Relative to the other algorithms, PDM2 shows poorer performance with RR Lyrae and eclipsing binaries. The STR algorithm also seems to fare worse with semiregular variables than with other pulsating types. However, the clearest differentiation between the algorithms comes with eclipsing variables. AOVMHW and CE are again the most successful and LS and GLS the least. If, however, we relax our accuracy criterion and also include (sub)harmonics of the true period then we find a significant improvement in LS and GLS relative to the other algorithms (see Fig.~\ref{p5comp}). Clearly, LS and GLS are the most susceptible of all the algorithms considered here to misidentifying a multiple of the period as the true value and this seems to be particularly the case with W UMa-type eclipsing binaries (hence the decline in performance at $\widetilde{\mu} \sim 3$.

\cite{dubath} report a correct period recovery rate of 91\% for LS/GLS depending on skewness value for non-eclipsing variable periods. The same approach finds a half period for 82\% of eclipsing variables. They note that better results are obtained with PDM with 38\% of the correct periods found and 38\% with half periods. However, for eclipsing variables, they adopt a strategy of only assigning a period once the object type has been assigned - doubling a LS-derived period for eclipsing binaries and ellipsoidal variables. Our results for LS/GLS certainly support this approach.

Finally, we note that \cite{drake12} find that VSX periods for RR Lyrae have an intrinsic error of $\sim 0.004$\% which equates to an accuracy cutoff of $\sim 10^{-4}$. This may contribute to the difference in recovery completeness seen in section 4.3.1 and 4.3.2 between cutoffs of $10^{-5}$ and $10^{-4}$. However, we have used a limit of $10^{-4}$ in this section for comparison and so any effect would be reduced.

\subsection{Variability}

\begin{figure*}
\centering
\caption{This shows the completeness fraction for the different period finding algorithms on the full combined data set in terms of the median absolute deviation (MAD) from the median of the light curve of the variable object. (a) gives almost the full range of MAD covered by the data set whilst (b) focuses on the smaller range covered by 97\% of the data. The three plots in each row again correspond to the different accuracy cutoffs: $10^{-5}$, $10^{-4}$, and $10^{-3} \, \mathrm{days}^{-1}$ over a 10-year baseline. The same symbols are used for each algorithm as in Fig.~\ref{res_mag} with the optimal frequency sampling used where relevant.}
\label{res_mad}
\includegraphics[width=7.0in]{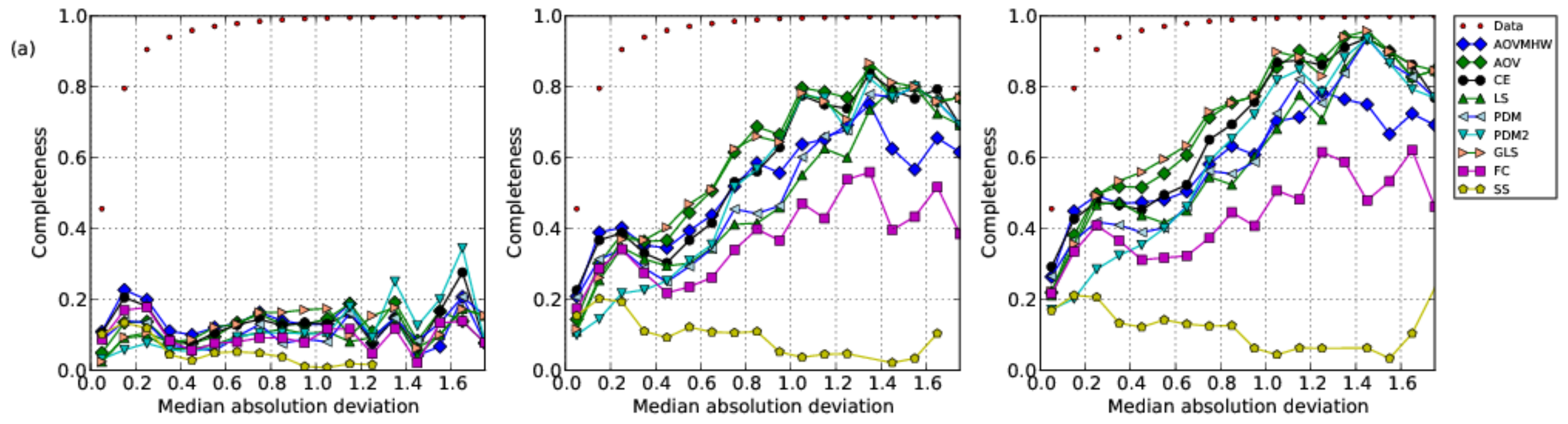}
\includegraphics[width=7.0in]{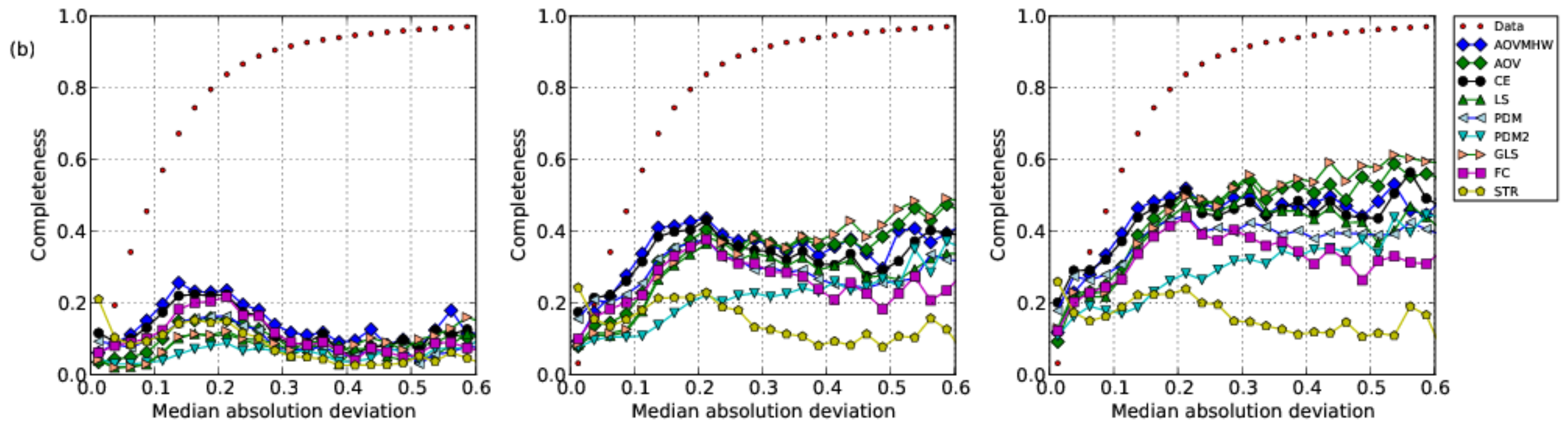}
\end{figure*}

The results in the previous section show the dependencies of the various algorithms on the specific object classes but it is also interesting to see whether there is any more general dependency on the variability of an object, i.e., it is easier to find the period of an object with strong variability than with weak. Note that the source of the variability here is in the physical nature of the star rather than in measurement errors which is covered by the dependency on the quality of the light curves (see section~\ref{quality}). Fig.~\ref{res_mad} shows the results for the algorithms as a function of median absolute deviation (from the median) - a more robust measure of the amplitude of variation than just the extrema values in a light curve. 

At the most conservative accuracy cutoff ($10^{-5}$ days$^{-1}$), there is essentially no dependency on object variability; however, with the other cutoffs, all the algorithms except STR show better performance with more variable objects. The objects with a correct period at the strictest cutoff tend to have more observations in their light curves so the periodic signal is already better sampled and the increased variability has no real effect. At the other cutoffs, those objects with a poorly sampled light curve due to fewer observations get a boost from larger amplitude variability which makes the periodic signal easier to detect by the algorithms. 

\begin{figure}
\centering
\caption{This shows the completeness fraction for the different period finding algorithms on the full combined data set in terms of the ratio of the median absolute deviation to log(SNR). An accuracy cutoff of $10^{-3}$ days$^{-1}$ was used. The same symbols are used for each algorithm as in Fig.~\ref{res_mag} with the optimal frequency sampling used where relevant.}
\label{noise}
\includegraphics[width=3.0in]{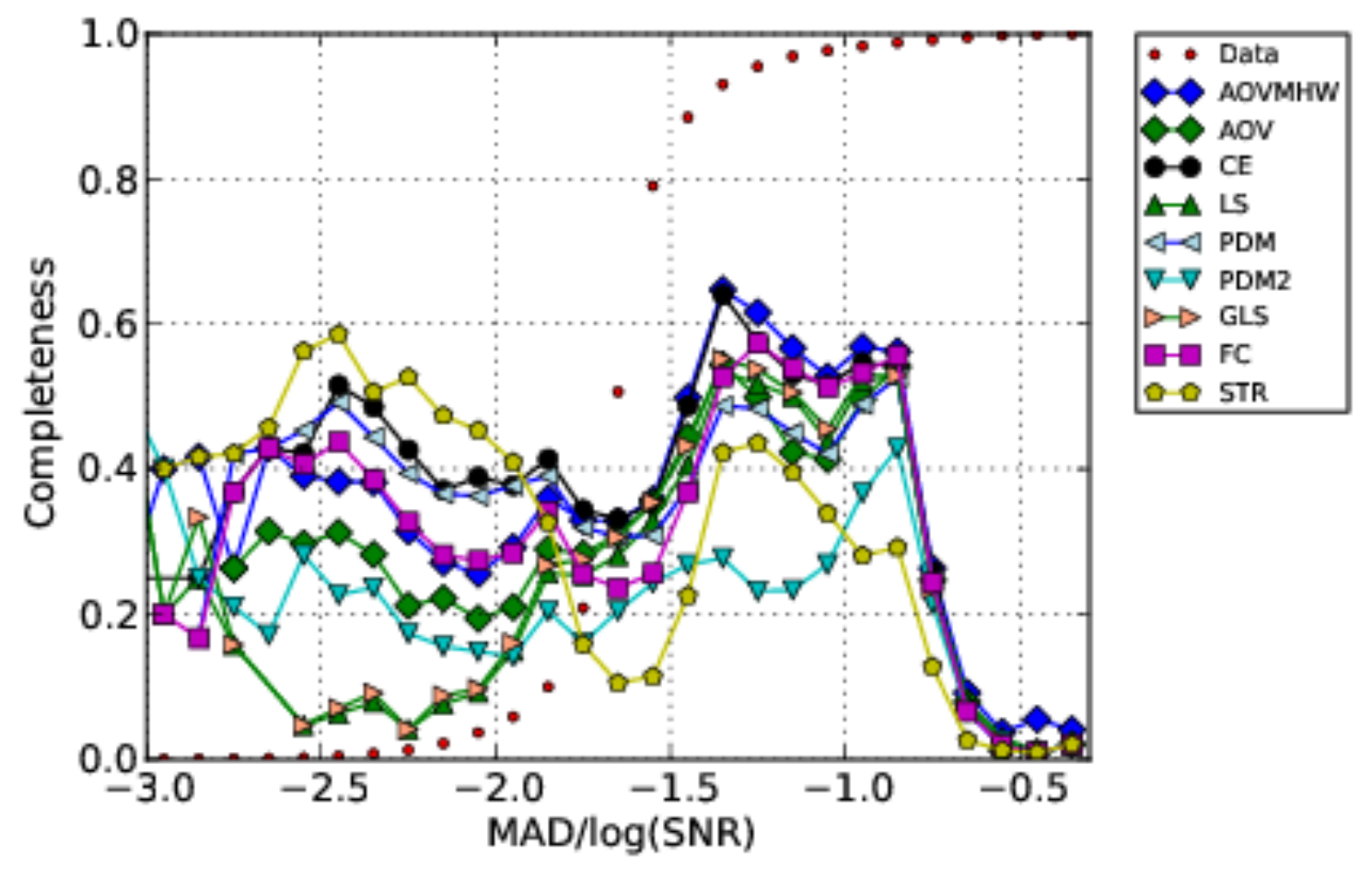}
\end{figure}

This behaviour is modulated by the noise characteristics of the light curves: objects with the same amplitude variability but different noise levels will have different period recovery accuracies. We can use the ratio MAD/log(SNR) as a proxy for the noise in the light curve and Fig.~\ref{noise} shows the recovery accuracy in terms of this quantity for the different algorithms on the combined data set. The structure in this plot is due to the individual contributions from the data sets with the initial peak at MAD/log(SNR) $\sim$ -2.5 from ASAS and that at $\sim$-1.2 from CRTS.

One difficulty with the low amplitude variability sources is that the phase errors could be substantial and yet we would not be able to visually recognize this, i.e.,  a correctly phased light curve and an incorrectly phased one are indistinguishable in the limit of vanishing variability - they both appear constant within observational error tolerance. This should particularly affect those object classes associated with low scale variability, such as small amplitude red giants or weak-line T Tauri objects. If we assume a photometric error of 0.05 mag then $\sim19$\% of objects have a MAD value less than this and could potentially have a misassigned period.

\subsection{Reliability}
\label{reliability}
\begin{figure}
\caption{This shows the distribution of the accuracies of AOVMHW (blue) and CE (red) periods for $\delta$ Scuti stars (P.2.6).}
\label{per_accuracy}
\includegraphics[width=3.4in]{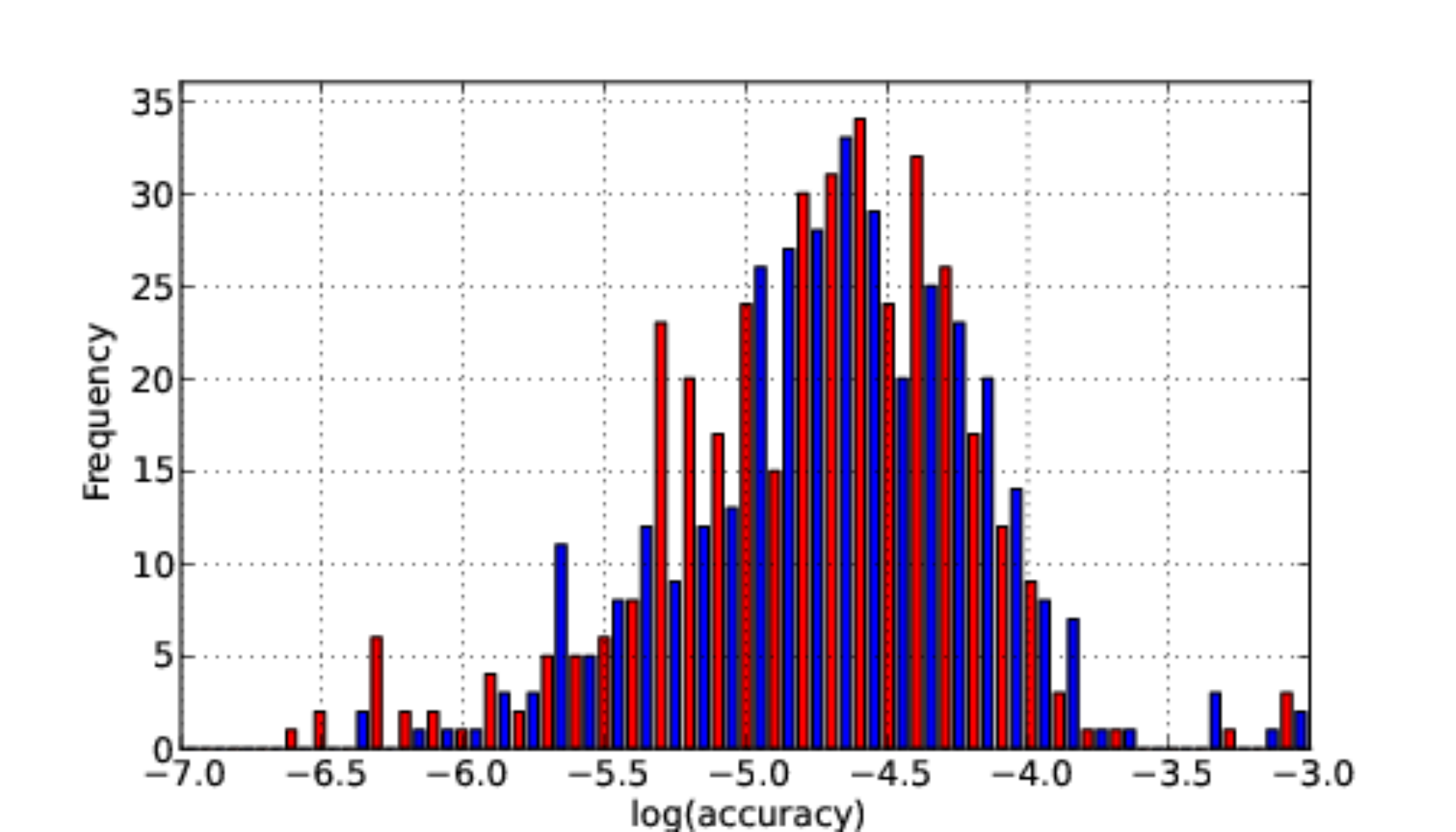}
\end{figure}

For each class in Table~\ref{types}, we have determined the most reliable method,  i.e., which method gives the most number of periods within an accuracy of $10^{-4} \mathrm{day}^{-1}$. This is a somewhat subjective measure since a method which finds a few highly accurate periods may be considered more reliable than one which gives a larger number of less accurate ones: when a correct answer is given, it will be very accurate but a larger number of workable periods might be more useful for a particular study. The accuracy limit of $10^{-4}$ reflects a suitable tradeoff between the two. Fig.~\ref{per_accuracy} shows the distribution of period accuracies for $\delta$ Scuti stars (DSCT, P.2.6) for AOVMHW and CE. Although the overall distributions are similar, CE provides slightly more accurate periods (below $10^{-5}$) whereas AOVMHW gives $\sim \!\! 10$\% more periods overall less than the $10^{-4} \mathrm{ day}^{-1}$ limit and so is the more ``reliable'' of the two.

\begin{figure*}
\caption{This shows the relative distribution of the accuracies of all the methods with the combined data set.}
\label{full_accuracy}
\includegraphics[width=7.1in]{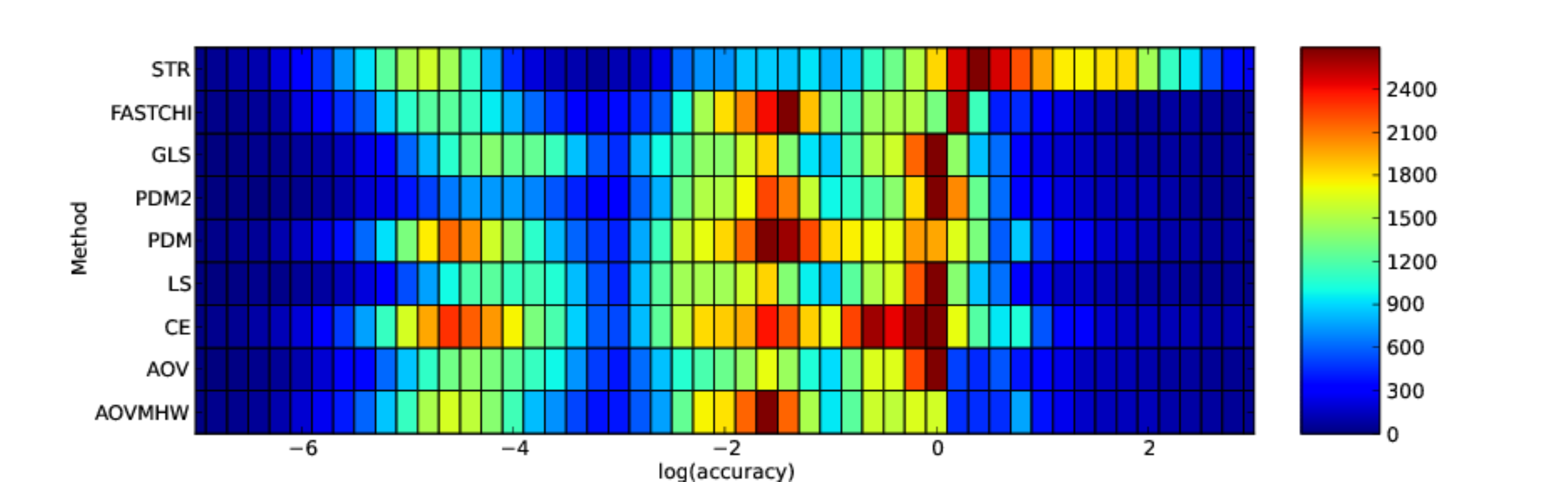}
\end{figure*}

Fig.~\ref{full_accuracy} shows the distribution of accuracies for all methods with the combined data set. This shows that CE and PDM both perform well below $10^{-4}$ along with AOVMHW. The poor reliability of PDM2 is also very clear. The peak at log(accuracy) $\sim$ 0 is largely due to methods finding half-periods for objects with periods around 1d. AOVMHW shows less susceptibility to this since it involves fitting higher harmonic orders.

\subsection{Ensemble method}

The accuracy of each of the individual algorithms is clearly dependent on observational factors such as the number of epochs in and the overall quality of a light curve as well as aspects natural to the source itself, such as the amplitude of variability and the actual object type. An ensemble approach, however, might serve to mitigate the effects of these dependencies and give a more robust and consistent result. While we reserve a full comparison of ensemble techniques to a forthcoming paper (Graham et al., in preparation), we will consider a simple approach here involving majority opinion.

Each light curve is associated with a set of period estimates, one for each algorithm considered. Within a set, we identify the largest subset of similar values, i.e., those which are within a specified tolerance of each other, and take the median value of this subset as the ensemble period estimate. Table~\ref{ensemble} gives the relative performance of AOVMHW, CE and GLS against the ensemble estimate for the three different accuracy cutoff levels used here. We have used the accuracy cutoff as the tolerance value for the three cases, although for a specific accuracy cutoff there is only $\sim$10\% variation in performance if a fixed tolerance of $10^{-5}$ is used. The results show that a simple ensemble approach does no better than the two strongest single algorithms. If we reduce the set of algorithms considered to just one of each type (AOVMHW, GLS, PDM, STR, CE, FC), we get similar results and just using the mean of the set performs poorly in comparison.

\begin{table}
\center
\caption{The relative performance of some of the algorithms compared against the ensemble majority opinion estimate in terms of total numbers of objects accurately measured at the various accuracy cutoffs.}
\label{ensemble}
\begin{tabular}{llll}
\hline
Algorithm & $10^{-5}$ & $10^{-4}$ & $10^{-3}$ \\ 
\hline
AOVMHW & 10804 & 20983 & 25402 \\
CE & 9980 & 20818 & 25746 \\
GLS & 4318  & 15230 & 22468 \\
Ensemble & 8452 & 18249 & 24516 \\
Mean & 1678 & 3075 & 8188 \\
\hline
\end{tabular}
\end{table}

The relative insensitivity of the ensemble result to the specific tolerance level used (10\% over 3 orders of magnitude) suggests that care should be taken when selecting values that are similar from multiple algorithms.

\subsection{Performance}
\label{performance}
The time taken by an algorithm to determine the period of a light curve is another important factor for large-scale automated analyses. Binning algorithms should show ${\cal O}(nN)$ behavior, where $n$ is the number of measurements and $N$ is the number of frequencies tested whereas FFT-based algorithms should exhibit an ${\cal O}(N \log N)$ dependency (\cite{fastchi}). GPU-based algorithms, at least for LS and GLS methods, should show ${\cal O}(nN)$ scaling tending to ${\cal O}(n^2)$ in the limit of large $n$ (\cite{gpuls}) Of course, the constant in front of the dependent terms in all cases will vary between algorithms and this can be the deciding factor too.

We have measured the computational time required by each algorithm to process each light curve in the MACHO data set with different frequency resolutions (spanning 0.1 to 10$^{-6}$ d$^{-1}$) on the same machine (an Apple iMac with a 2.8 GHz Intel Core i7 CPU and 8 GB 1333 MHz DDR3 memory running Mac OS X 10.7.4 and AMD Radeon HD 6770M with 512MB for the GPU algorithms) and then performed a linear regression fit to the time taken in seconds as a function of the expected behavior. Note that the frequency resolution cannot be set as an argument for SS and CKP and so we just estimate $N$ for these algorithms based on their documentation. We find that the binned algorithms are better described by a ${\cal O}(n^xN^y)$ relationship than a strict ${\cal O}(nN)$ one but the FFT-based algorithms agree well with the expected $N \log N$ dependency. The GPU-based algorithms (LS, GLS) show an essentially constant timing behaviour to $N = 10^{5}$ and then transition to ${\cal O}(nN)$. Unfortunately we do not have sufficiently large values of $n$ relative to $N$ to show the asymptotic scaling. The constant term is an implementation artifact, attributable to memory overheads in transferring data between the CPU and GPU, and only for $N > 10^5$ does the GPU computation begin to take a discernible amount of time. 

Table~\ref{timings} gives the details of the regression fits to the respective behaviours. Whilst the absolute performance of the algorithms will depend on the hardware used (CPU speed, memory configuration, etc.), 
the constant values, $(A)$, can give a reasonable indication of their relative speeds. For example, there is a clear factor of $\sim 15$ in performance time between the two versions of AOV - the faster just relying on binning and the slower on model fitting. The two GPU-based algorithms also show a slight difference with GLS being slightly slower as it involves slightly more trigonometric function calls. Note that the intercept $(c)$ for these indicates the memory overhead time and so for small values of $N$ these are not particularly performant. 

In fact, we can estimate the minimum approximate time taken by the relative algorithms to determine the period for any light curve and frequency sampling. For a light curve consisting of $n$ observations covering a timespan $T$ and with a fastest timescale of interest, $\delta t$, the maximum frequency $\nu_{max} \ga 1/2 \delta t$ and frequency sampling $\delta \nu \la 1 / T$. The minimum number of frequencies to test is then $N_{min} \simeq T / 2 \delta t$.

\begin{figure*}
\caption{This shows the distribution of the accuracies and timings in seconds for the algorithms considered in this analysis applied to the MACHO data set with $n = 966$ and $N_{min} = 6803$ for (a) exact periods and (b) including period harmonics. The three plots in each row again correspond to the different accuracy cutoffs: $10^{-5}$, $10^{-4}$, and $10^{-3} \, \mathrm{days}^{-1}$ over a 10-year baseline.}
\label{tsc_accutime}
\includegraphics[width=7.1in]{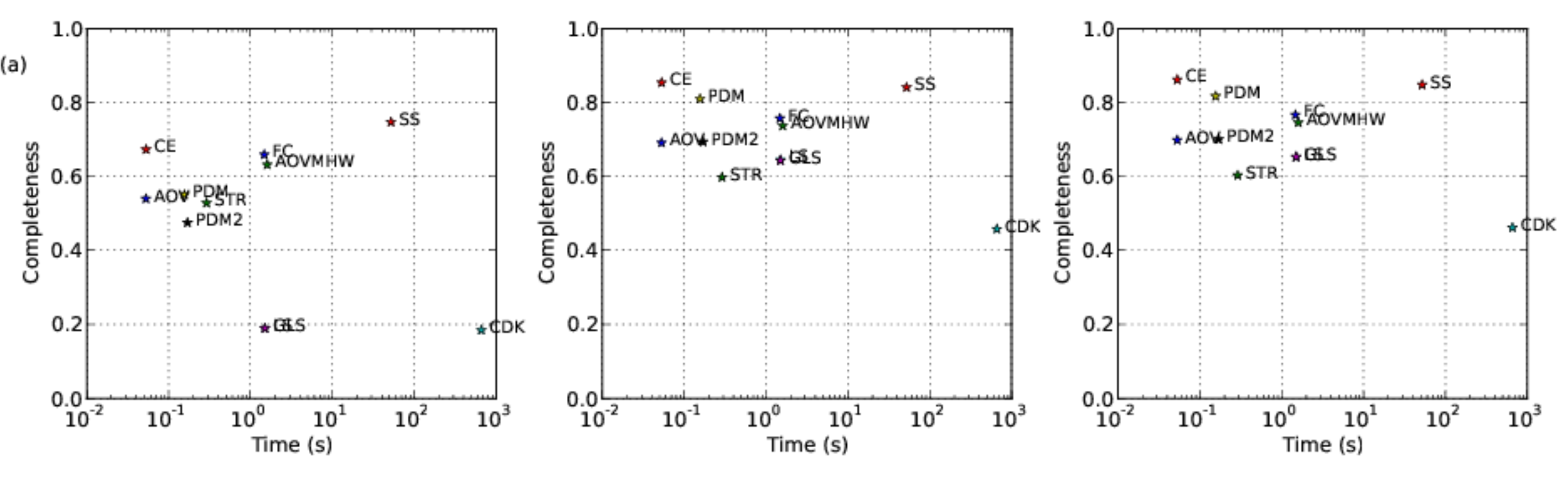}
\includegraphics[width=7.1in]{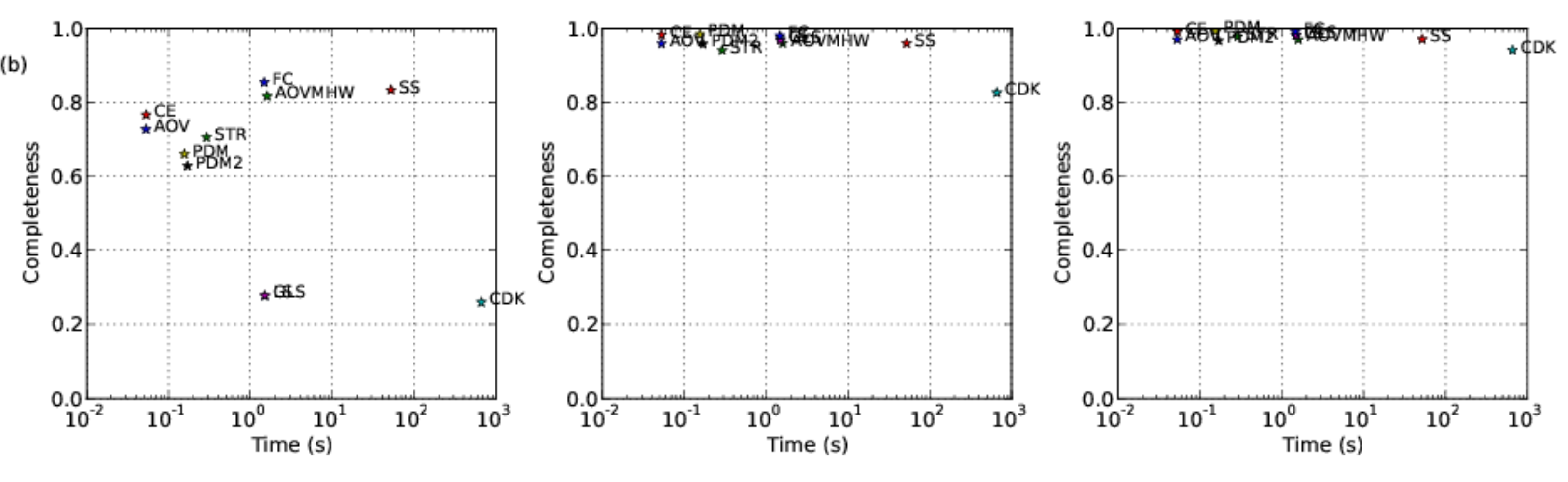}
\end{figure*}

The MACHO data set consists of RR Lyrae, Cepheids and eclipsing binaries and so a realistic fastest timescale of interest is $\delta t = 0.2$ days. Taking the median timespan ($T = 2720.881$d) and number of observations ($n = 966$), we can calculate a representative minimum time for each algorithm as an indicator of performance. 
Fig.~\ref{tsc_accutime} shows the accuracies vs. performance for the algorithms applied to the MACHO data set, for both exact periods and harmonics.

\begin{figure*}
\caption{This shows the distribution of the accuracies and timings in seconds for the algorithms applied to the regular periodic variables data set with $n=347$ and $N_{min}=648425$ for (a) exact periods and (b) including period harmonics.  The three plots in each row again correspond to the different accuracy cutoffs: $10^{-5}$, $10^{-4}$, and $10^{-3} \, \mathrm{days}^{-1}$ over a 10-year baseline. }
\label{goodper_accutime}
\includegraphics[width=7.1in]{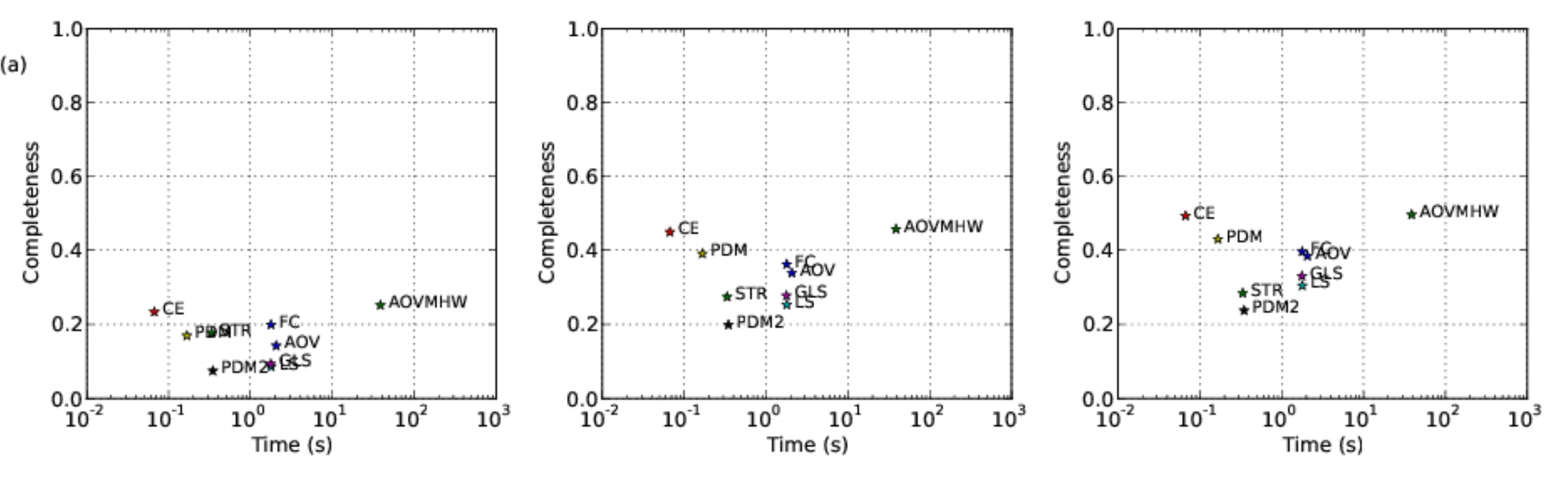}
\includegraphics[width=7.1in]{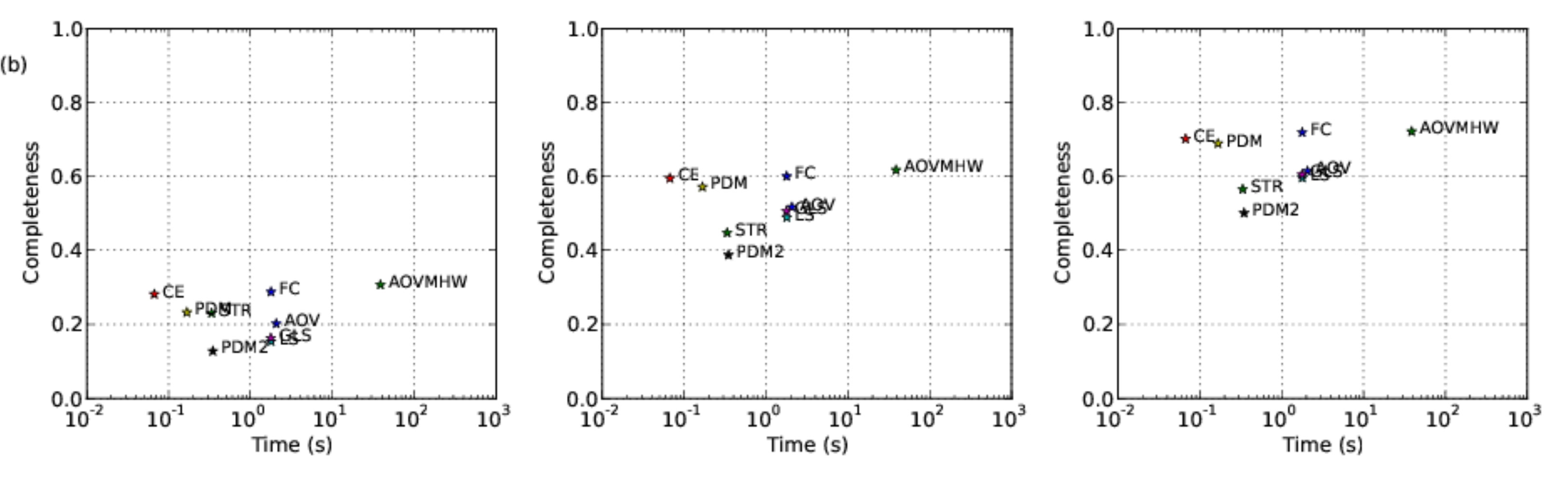}
\end{figure*}

We have also considered a set of regular periodic variables drawn from all three surveys, consisting of all objects in the following classes or class families: T Tauri (P.1.3.3.3), RS Can Ven (P.1.5), Beta Cepheid (P.2.2), Cepheid (P.2.3), W Vir (P.2.4), Delta Cep (P.2.5), Delta Scuti (P.2.6), Mira (P.2.8), RR Lyrae (P.2.10), rotating (P.3), and eclipsing binary (P.5.1). This has a total of 40550 members with a median timespan of 2593.97 days and a median number of observations of 347. Fig.~\ref{goodper_accutime} shows the accuracies vs. timings for the algorithms when applied to this data set, assuming a fastest timescale of interest of $\delta t = 0.02$ (for Delta Scuti objects) and a frequency oversampling of 10.  In both sets of plots, the ideal algorithm will be closest to the top left of the plot, i.e., high completeness and low timing slope (fast). We identify this as the CE method.


\begin{table*}
\center
\caption{The parameters of the regression fits to the timings of the various algorithms in seconds as a function of number of observations in the light curve, $n$, and the number of trial frequencies tested, $N$: $t=An^{x}N^{y} + c$ or $t=AN\log N + c$ respectively. An asterisk indicates a GPU-based algorithm. Note that the GLS and LS fits are only valid for $N>10^5$, otherwise a constant value of 1.5s should be assumed.}
\label{timings}
\begin{tabular}{llllll}
\hline
& Algorithm & $\log A$ & $x$ & $y$& $c$ \\
\hline
${\cal O}(n^xN^y)$ & AOV & -7.939 & 0.987 & 0.989 & -0.010 \\ 
& AOVMHW & -6.754 & 0.997 & 0.998 & 0.480 \\ 
& PDM& -9.446 & 0.686 & 0.990 & 0.156\\ 
& PDM2 & -5.067 & 0.948 & 0.376 & 0.010\\ 
& STR & -9.846 & 1.073 & 0.995 & 0.289 \\ 
& CE & -8.921 & 0.600 & 0.955 & 0.053 \\ 
& SS & -1.293 & 1.007 & 0.0 & 0.436\\ 
& CKP & -3.166 & 2.009 & 0.0 & -16.3 \\
& LS* & -2.732 & -0.007 & 0.513 & 0.078 \\ 
& GLS* & -2.793 & -0.007 & 0.523 & 0.088 \\ 
\hline
${\cal O}(N \log N)$ & FC & -7.085 & - & - & 1.472\\ 
\hline
\end{tabular}
\end{table*}

\section{Discussion}
The results in the previous section show that at best period finding algorithms can recover the period of a regularly periodic object with a reasonable degree of accuracy (an equivalent phase offset between $10^{-3}$ and $10^{-4}$ days$^{-1}$, say, over a 10-year baseline) in only about 50\% of cases. If one is only interested in detecting periodic behaviour, i.e. the period or a (sub)harmonic, then rates of $\sim \! 70$\% are achievable. For objects which do not show simple periodicity, i.e., they are semi-periodic, quasi-periodic or multi-periodic, the situation is broadly much worse, typically only around 10--20\% of cases. Of course, the fundamental assumption underlying this analysis is that the quoted period is correct. We have been careful, however, to use data sets where all the light curves have been inspected and the periods confirmed visually.

It should be noted that many of the algorithms score very highly when tested on simulated periodic signals,  typically sinusoids with Gaussian noise; the problem seems to come with real data. For many objects, quoted periods would have originally been determined from a small number of observations over a short time baseline. The advent of large-scale synoptic sky surveys means that hundreds of observations over baselines of 5 to 10 years are now readily available and future projects such as LSST will extend this to baselines of a couple of decades. The digitization of the Harvard plate library (DASCH\footnote{http://hea-www.harvard.edu/DASCH}) offers multidecade baselines for many objects as do other similar historical collections. At the other end of the scale, exoplanet searches and space astroseismology projects, such as SuperWASP, Kepler and CoRoT, are providing (very) high resolution samplings of a few periodic cycles over periods of days and months.

This wealth of new information allows the long-term stability of periods to be examined as well as intra-/intercycle variations and is now suggesting that, even for astrophysical objects exhibiting periodicity, a single value is not capable of characterizing their temporal behaviour. Kepler results have shown that 60\% of dwarf stars are more variable than the Sun and probably pulsating variables (\cite{mcquillan}). RR Lyrae are one of the most populous of pulsating variables and employed as standard candles in studies of galactic structure. However, it has long been known that $\sim10\%$ exhibit a long-term, generally quasi-periodic modulation of widely varying strength known as the Blazhko effect. Studies of variable stars in M3 now show that about a third of RR Lyrae display Blazhko behaviour and the discovery of small amplitude cycle-to-cycle modulations of RRabs (\cite{szabo}), in addition to Blazhko effects, cautions that large surveys may have seriously underestimated the number of modulated RR Lyrae stars.

For other populous classes, the situation is equally as complicated with many types of variables showing cyclic period changes over multidecade baselines, such as close binary systems (\cite{zavala}) and long period variables (\cite{lebzelter}). \cite{sterken96} note that a subgroup of semi-regular variables show very clear double periods. In some cases the longer period may be due to orbital effects indicating that the star is in a binary system. Other semi-regular variables apparently show multiperiodicity (e.g., \cite{kerschbaum}), but in general it is not clear whether these stars are truly multiperiodic, chaotic or both, although the actual existence of irregular, i.e., non-periodic, variables among red giants is in dispute (\cite{obbrugger}). 


The traditional approach to characterizing periodicity variation is the O-C diagram (e.g., \cite{sterken}) which tracks the evolution of the time of appearance of a feature (say the light curve maximum) relative to the corresponding multiples of the period. The functional form of the period change ($dp / dt$) determined from it can be used in principle to infer the physics of the situation, e.g, a steadily increasing pulsation period implies an expanding star; however, stochastic evolution, e.g., the mean period follows a random walk, can produce equivalent effects in the O-C diagram and distinguishing between the two is an area of active research (\cite{koen07}). However, the method cannot be applied to long-period pulsating variables where the intrinsic scatter of the period is usually comparable to the experimental error in the period determination (\cite{lombard92}). It also has issues with multiperiodic light curves and those with strong modulation. 
     
Alternate approaches rely on techniques from communication and signal processing theory, e.g.,  wavelets (\cite{foster,blackman}),  carrier signals (\cite{pelt}) and other time frequency analysis methods. Though these can be very powerful, they are complicated and it is difficult to distill the results down to a single useful characterizing feature, akin to the period. It is possible that the first derivative of the period as a measure of  periodicity variation or the (largest) Lyapunov exponent to describe the degree of chaos in a time series (\cite{wolf}) may be suitable; however, further discussion of these is outside the scope of this paper.

Another issue potentially affecting the results in this paper is that of object misclassification or class uncertainty. \cite{dubath} only assign a period to eclipsing binaries and ellipsoidal variables once the object type has been determined (to mitigate the half-period issue with these objects). Our results support this as 
a viable strategy for the LS/GLS algorithms: the improvement seen on our combined data set is $\sim 4\%$ recovery to $\sim 50\%$ whereas other algorithms show a significant drop. The biggest source of error, however, will be those objects that have been misidentified as eclipsing variables, although this could be mitigated by a high classification accuracy for eclipsing binaries. Whilst a detailed discussion on object classification is beyond the scope of this paper, we will note a few points. 

The MACC classes that we employ for the ASAS data use a probabilistically determined 28-term scheme whilst the original ACVS classifications for the same objects used 439 different categories (different combinations and permutations of a set of about 20 terms), although 60\% of objects were classified as  `MISC'. One of the hardest classes to distinguish between is RR Lyrae with fundamental overtones (RRC) and W UMas (EC) and the effect of misidentification would be that a half-period (for a W UMa) is reported as the true period (for a RRC). 12\% of MACC W UMas are considered to be RRCs by ACVS and about 10\% vice versa, although in only a handful of cases are the probabilities of both classes within 5\%. The MACHO data set shows a similar level of misclassification ($\sim 10\%$) between the provided object type (RR) and the MACHO assigned class (eclipsing binary). We therefore estimate that there may be $\sim 10\%$ error in the class-based results arising from misassigned object types. Note, however, that if data from more than one band is available then these types can be better distinguished with PCA (\cite{suveges}).

It is also possible to use additional data to check the classifications, e.g., T Tauri/HAEBE stars and any massive star classes should be near the plane. WTTS objects in the ASAS data set do not correspond very well with the plane or nearby star forming regions casting doubt on the reliability of these classifications (Feigelson, private communication). We have also compared the reported periods for objects against the expected period ranges for their class drawn from \cite{deboss07} and object definitions in VSX. We find that of the 41299 objects in the combined data set for which we have ranges, 40052 have periods which lie within the expected class ranges.
This is certainly well within the $\sim 10\%$ misclassification error we have estimated and suggests that class uncertainty is not a significant issue in this analysis. We note, however, that coupling classification and period finding may produce more accurate results, e.g., \cite{rimoldini} finds improved classifications using a weighting scheme based on the period-folded light curve.

\section{Conclusions}
In this paper we have analyzed the performance and dependencies of the most popular period finding algorithms against a comprehensive set of light curves. We find that:

\begin{itemize}
\item all methods are dependent on the quality of the light curve and show a decline in period recovery with lower quality light curves as a consequence of fewer observations, fainter magnitudes and/or noisier data and an increase in period recovery with higher object variability;
\item all algorithms are stable with a minimum bin occupancy of ~10 (assuming $\Delta \phi = 0.1$);
\item a bimodal observing strategy consisting of pairs (or more) of short $\delta t$ observations per night and normal repeat visits is better than single observations with normal repeats;
\item a minimum frequency step of $\delta \nu = 0.0001$ is sufficient;
\item the algorithms work best with pulsating and eclipsing variable classes;
\item straightforward ensemble methods show no improvement over single algorithms. 
\end{itemize}

We also confirm that LS/GLS are strongly effected by the half-period issue for eclipsing binaries and find that 
PDM2 has issues with irregular sampling of light curves and that AOVMHW and CE work well at bright magnitudes (containing saturated values). Finally, in terms of overall performance factors considered here - greatest period recovery and time - CE is the best algorithm with AOV and PDM viable alternatives.

New and better techniques may be proposed that change the findings of this analysis. To keep track of these, we intend to maintain an online version of this work, updating it as appropriate. If anyone has an algorithm that they would like to see included then they should get in touch with us.

\section*{Acknowledgments}

We thank the referee, Lorenzo Rimoldini, for their meticulous reading of the paper and useful comments. We thank Eric Feigelson for useful discussions on this work and the various providers of software.

This work was supported in part by the NSF grants AST-0909182 and IIS-1118041, by the W. M. Keck Institute for Space Studies, and by the U.S. Virtual Astronomical Observatory, itself supported by the NSF grant AST-0834235. 

This research has made use of data obtained from or software provided by the US Virtual Astronomical Observatory, which is sponsored by the National Science Foundation and the National Aeronautics and Space Administration.

This research has made use of the SIMBAD database, operated at CDS, Strasbourg, France and the International Variable Star Index (VSX) database, operated at AAVSO, Cambridge, Massachusetts, USA.

\appendix

\section[]{GPU versions of Lomb-Scargle algorithms}
\label{appa}
Graphics processing units (GPUs) offer a significant performance improvement for parallelizable algorithms (see \cite{gpu} for a review of their potential for astronomy).  \cite{gpuls} provides a Lomb-Scargle periodogram code implemented within NVIDIA's CUDA framework. We have ported this to OpenCL (Open Computing Language) which provides a platform-neutral manner to program devices such as multicore CPUs and GPUs at a slight performance expense. This allows us to run the code on non-NVIDIA devices, such as AMD Radeon GPUs.  We have also implemented an OpenCL version of the generalized Lomb-Scargle periodogram (\cite{gls}). Details of both are given below.

\subsection{Porting CUDA Lomb-Scargle to OpenCL}
Porting the CULSP computation kernel essentially consists of just three steps.

\subsubsection{Rewriting the kernel signature}
The kernel signature under CUDA is:

\begin{verbatim}
__global__ void
__launch_bounds__(BLOCK_SIZE)
culsp_kernel(float *d_t, float *d_X, float *d_P, 
   float df, int N_t) {
\end{verbatim}

Under OpenCL, this becomes:

\begin{verbatim}
__kernel void culsp_kernel(__global float *d_t, 	
  __global float *d_X, __global float *d_P, 
  float df, int N_t) {
\end{verbatim}

\subsubsection{Thread management}
OpenCL has global commands for addressing threads so \texttt{blockIDx.x} is given by \texttt{get\_group\_id(0)} and \texttt{threadIdx.x} by \texttt{get\_local\_id(0)}. Synching threads within a block, \texttt{\_syncthreads}, is replaced with \texttt{barrier(CLK\_LOCAL\_MEM\_FENCE)}. Shared memory is also allocated with a \texttt{\_\_local} keyword instead of \texttt{\_\_shared\_\_}.

\subsubsection{Intrinsic function calls}
The OpenCL library has slightly different versions of certain functions to CUDA. \texttt{rintf} is \texttt{rint} under OpenCL and the CUDA function call \texttt{\_\_sincosf(TWOPI*ft, \&s, \&c)} becomes a variable assignment: \texttt{s = sincos(TWOPI*ft, \&c)}.

\subsection{An OpenCL Generalized Lomb-Scargle kernel}
\begin{figure*}
\centering
\caption{This gives an OpenCL computation kernel for the generalized Lomb-Scargle periodogram.}
\label{glsker}
\includegraphics[height=9.0in]{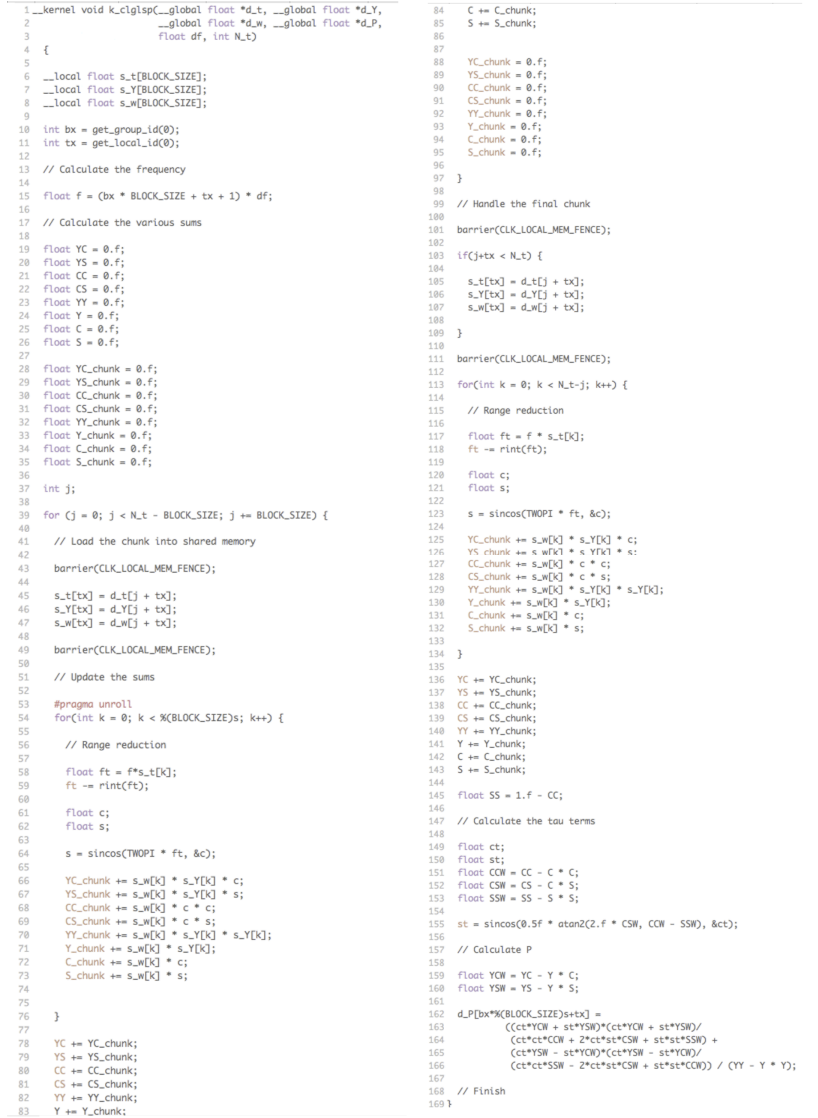}
\end{figure*}

As noted in \cite{gpuls}, the expressions derived in \cite{gls} to calculate the generalized Lomb-Scargle (GLS) periodogram are very similar in form to those used in the CUDA LS kernel. It is therefore straightforward to construct a GPU kernel for the GLS (see Fig.~\ref{glsker}). 

\bsp

\label{lastpage}

\end{document}